 \documentclass[final,5p,times]{elsarticle}

\usepackage{amssymb}

\usepackage{lineno}
\usepackage{color}
\usepackage{floatrow}

\journal{Progress in Nuclear Magnetic Resonance Spectroscopy}

\begin{document}
\begin{frontmatter}



\title{Journey to the Centre of the Earth: Jule Vernes' dream in the laboratory from an NMR perspective}


\author{Thomas Meier}

\address{Bayerisches Geoinstitut, Universität Bayreuth,Universitätsstraße 30, D-95447 Bayreuth, Germany}

\begin{abstract}

High pressure nuclear magnetic resonance is among the most challenging fields of research for every NMR spectroscopist due to inherently low signal intensities, inaccessible and ultra-small samples, and overall extremely harsh conditions in the sample cavity of modern high pressure vessels.
This review aims to provide a comprehensive overview of the topic of high pressure research and its fairly young and brief relationship with NMR.

\end{abstract}


\end{frontmatter}
\tableofcontents


\section{An unexpected journey for an NMR spectroscopist, a motivation}
\label{intro}

\begin{center}
"Pressure. To most people the word brings to mind the stress of our lives in a time of economic crisis. Yet to many scientists, pressure means something very different; it is an idea filled with wonder and power \--- a phenomenon unlike anything else we know. Pressure shapes the stars and planets, forges the continents and oceans, and influences our lives every moment of every day."
\end{center}
These are the opening lines of Robert Hazen's intriguing monograph "The New Alchemists", in which the author describes the early days of modern high pressure science, and its evolution towards one of the most challenging and fascinating research branches today\cite{Hazen1993}. It is precisely because of this challenge and uniqueness of high pressure research that we need to take a closer look at the methods available today, which allow us to recreate extreme conditions in the tiniest of spaces inside a laboratory. Thus, in this review, we will metaphorically retrace the footsteps of Jules Vernes' Professor Lidenbrock as he descends beneath the surface of the Earth, towards a strange and unknown and yet utterly enticing world - which we now know to be under extreme pressure. \\
But why investigate pressure with NMR?\\
While this simple question is frequently raised in an NMR environment, it is not that easy to answer. 
\begin{figure}[h]
  \includegraphics[scale=.2]{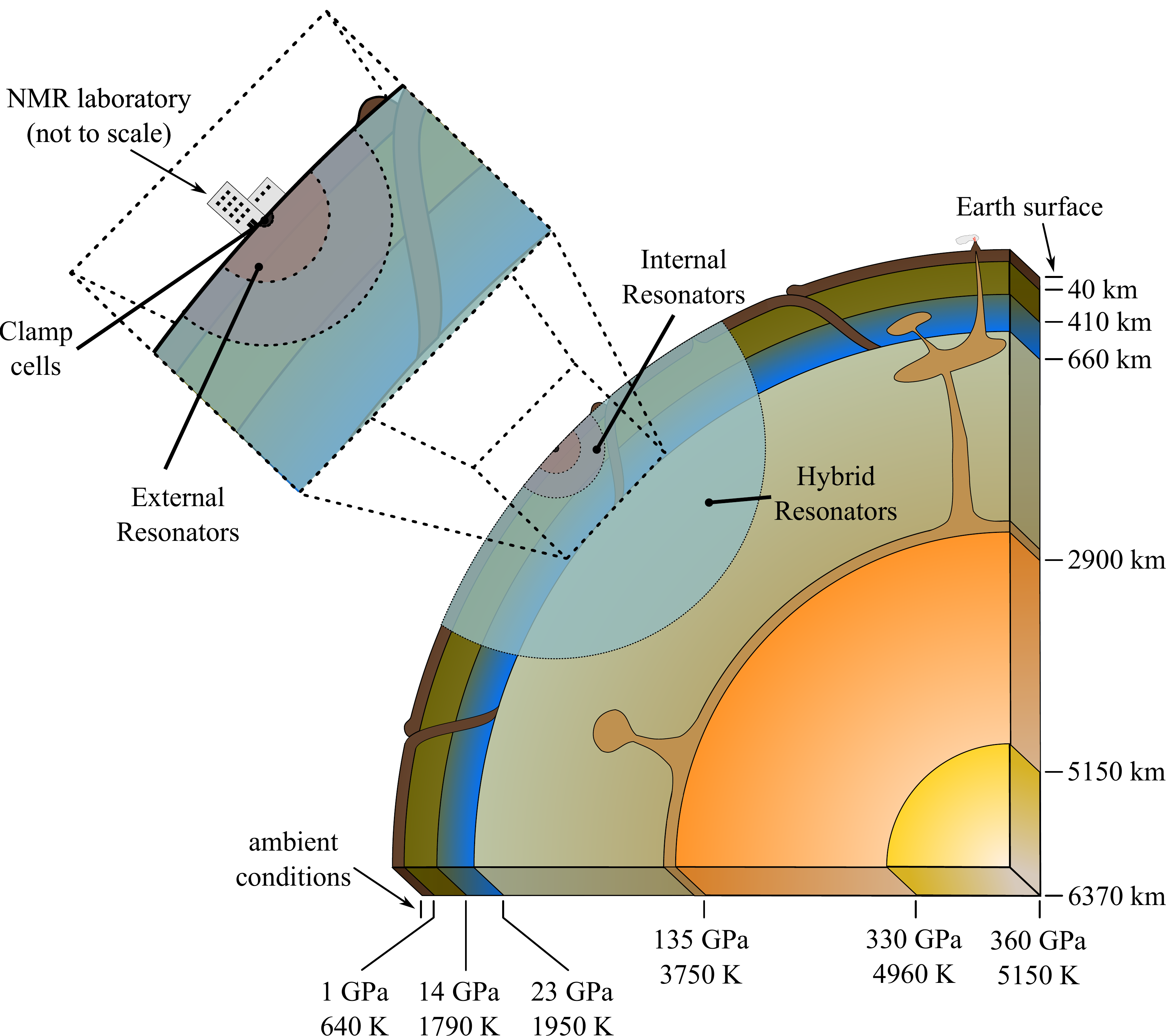}%
 \caption{Even on our planet Earth, pressure ranges over about seven orders of magnitude from its surface, at one bar of atmospherical pressure, to the gravitational centre with about 360 GPa. The schematic picture shows the inner structure of the Earth together with a pressure-temperature scale and corresponding depths. Up to now, NMR techniques are able to mimic the extreme conditions of the inner mantle \--- disregarding elevated temperatures for now\--- corresponding to a depth of about 2000 km.   
 \label{fig1}}
 \end{figure}
As far as NMR is concerned, the most typical ways to alter a given system are by altering its chemical composition, e.g. through doping, increasing or decreasing temperature, or using high or low magnetic fields. Pressure modifying vessels, on the other hand, are only rarely employed in NMR research, owing to substantial technical difficulties associated with the design of such apparatuses; their demand for radio-frequency (RF) resonators, which require most NMR spectroscopists to look over the rim of their tea cup; as well as their inherent inability to use modern high sensitivity and line narrowing methods like MAS or DNP.\\  Nevertheless, in condensed matter systems, where the interatomic bond strength easily exceeds several eV, utilisation of immense pressures is needed to induce significant structural or electronic changes. Other ways of increasing energy density would require tremendous resources, or else are simply extremely impractical. For example, the application of an external magnetic field of 50 T would only correspond to the application of about 1 GPa (10.000 atmospheres)\footnote{The energy density of a magnetic field is given by $\rho_{B}=B^2/2\mu_0$, yielding $9.94\cdot 10^8 Jm^{-3}$ for 50 T which is roughly 1 GPa (pressure has the same dimension as energy density)}; a pressure which is easily reached in modern diamond anvil cells.\\
If we would start to compare these pressures with the geothermal gradient of our planet Earth, we would have started our hypothetical journey at the bottom of the Mariana's trench,  corresponding to about 1 kbar or 100 MPa of pressure exerted by the 11 km high water pillar above us. Reaching deeper into Earth's interior, say to a depth of about 200 km in the middle of the upper mantle, the experienced pressure would be similar to the contact pressure of the Eiffel tower turned up side down and balanced on its tip, which would only be 10 GPa (about 100.000 bar). At one mega-bar, or 100 GPa, we would already have "dived" down half way through the lower mantle, and at a crushing 350 GPa our journey would come to a violent end upon reaching Earths' gravitational centre, a solid ball primarily comprised of iron and nickel. Clearly, Professor Lidenbrock and his companions would have come to a literally crushing end as well, had they succeeded in their quest.\\
In modern high pressure laboratories, of course, Jules Vernes' fictional journey takes a somewhat different path. Here, pressures are generated by the application of pressure sustaining vessels, which are often composed of a movable piston and an enclosed pressure chamber\cite{Bridgman1952a}. Especially with the invention of the diamond anvil cell by Charles Weir\cite{Weir1959} and Alvin van Valkenburg\cite{Piermarini1993} in 1959, high pressure science took up momentum and became an integral part of contemporary chemistry, biochemistry, physics, and geophysics\cite{Grochala2007, Bassett2009, Hemley2010}. 
As the technique evolved with higher quality diamonds of increasingly complex geometries, and harder materials used for the pressure vessels, the range of applicable pressures rapidly increased well into the megabar regime (1 Mbar = 1 million atmospheres)\cite{Mao1985, Hazen1993}, mimicking the extreme conditions at the centre of the Earth. Recently, record pressures of up to 1 TPa were achieved in double-stage diamond anvil cells\cite{Dubrovinskaia2016}.\\
This review's aim is threefold. Firstly, readers not familiar with this rather exotic application of magnetic resonance\footnote{I am omitting the specification of "nuclear" here on purpose, because the techniques presented are also applicable for pulsed ESR methods.} should gain a general impression of the "nuts and bolts" approach associated with these experimental set-ups. Secondly, I will review NMR experiments obtained using micro-coil and magnetic flux tailoring techniques, which are capable of reaching pressures from between 1 GPa and close to 1 Mbar. Closely related to this is a critical overview of some methodological difficulties arising within certain experiments, which can complicate data analysis, or might even lead to false interpretations.\\

 
\section{From Psi over Torr to kbar. The Low pressure regime as the playground for bio-chemistry and life-sciences.}
\label{lowp}

\begin{figure}[h]
  \includegraphics[scale=.25]{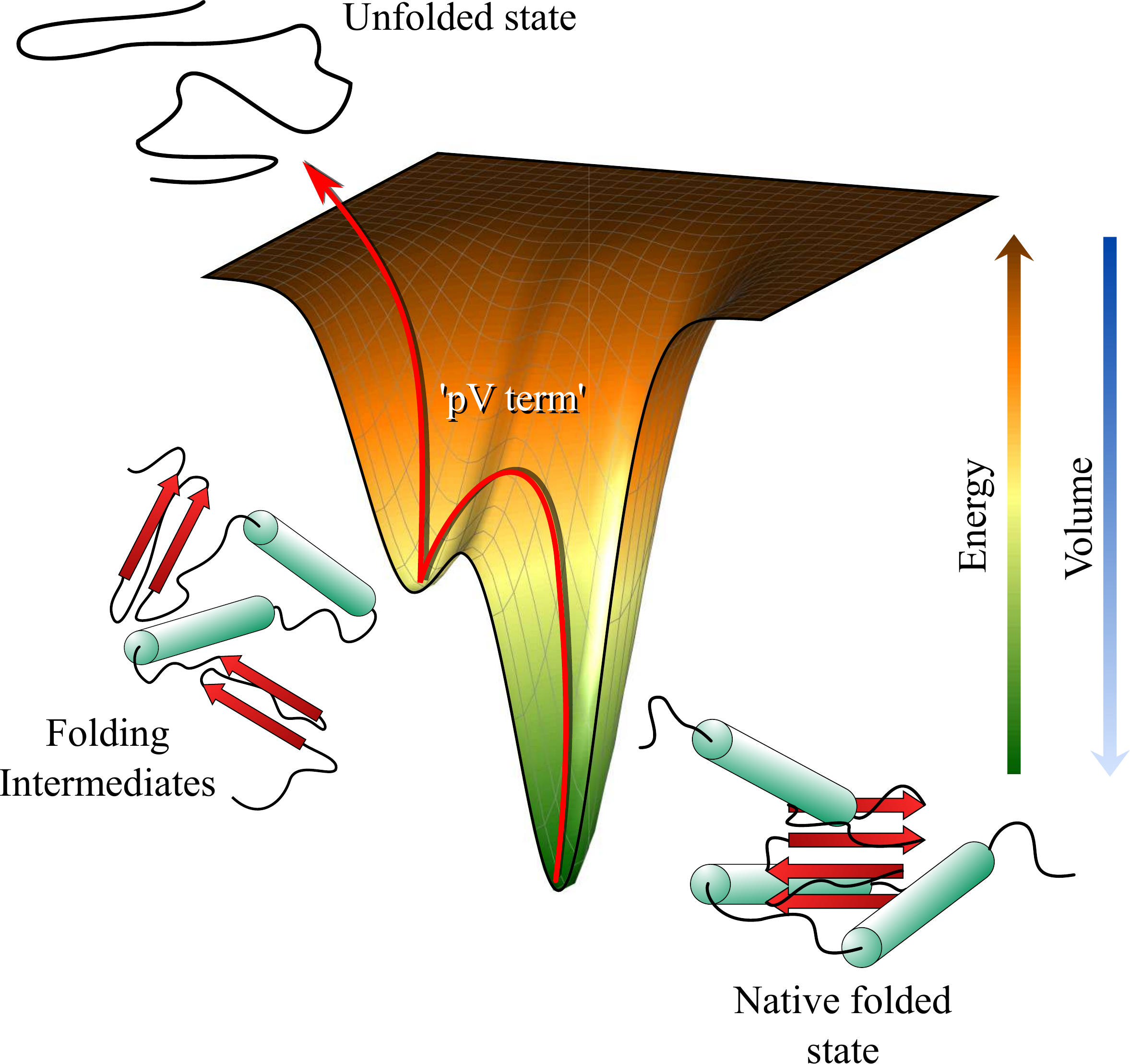}%
 \caption{Schematic representation of the protein-volume-theorem of an arbitrary protein in an arbitrary energy landscape. The native folded molecules often occupy the highest molar volume, whereas a decrease in the overall volume leads to a destabilisation of the protein structure.   
 \label{fig2}}
 \end{figure}

Let us begin our journey with a pressure regime ranging over almost four orders of magnitude from ambient conditions to about 1 GPa. This is the realm of high pressure Bio-NMR, were one of the more commonly used and better known high pressure NMR set-ups is used. This approach uses so-called clamp cells, which are basically comprised of a movable piston exerting the pressure on a sample volume often as big as 100 $\mu l$ . \\
As this pressure range is still below solidification transitions of most liquid buffer media, it has been proven to be an ideal tool investigating pressure driven protein folding and unfolding dynamics in liquids\cite{Kitahara2003, Li2006} under increasing compressional stages. Figure \ref{fig2} schematically demonstrates the famous protein volume theorem first proposed by Kitahara et al.\cite{Kitahara2003} and
Li et al.\cite{Li2006}.

As this field of high pressure NMR research is extremely extensive and already very well studied, I would like to draw the attention of the reader to the comprehensive review articles from Jonas\cite{Jonas1994}, Ballard\cite{Ballard1997}, and Roche et al.\cite{Roche2017}.

\section{Within the Upper Mantle: First Endeavours with DACs}
\label{midp}

Up to this point in the pressure-temperature landscape, experimental conditions mimicking the pressure at the bottom at the Marianas trench, and some kilometers deeper in the Earth's crust, could easily be achieved without the application of diamond anvil cells. However, conditions beyond 1 GPa demand a much more powerful device.\\
In this sense, the DAC turned out to be one of the most versatile pressure generating vessels, as it enables the experimenter to not only reach very high static pressures, but also provides him with an astonishing variability of set-ups which can be adopted to a plethora of different experimental environments.\\
Nevertheless, performing NMR experiments in a DAC can at best be considered problematic due to the following reasons. \\
1) The available sample volume in a DAC is often several orders of magnitude smaller than in a standard NMR experiment. The reason behind this is obvious: We can either generate high pressures by applying a huge force on a sample of some dozens of mm$^3$, which becomes exceedingly unpractical as we reach pressures above, say, 2 or 3 GPa. Also, these such so-called "large volume" presses are typically fairly large \cite{Liebermann2011} (several meters in height, for example), thus an application in a superconducting NMR magnet would be out of the question. The other possibility is, of course, to reduce the size of the pressurising area. In a DAC, typical culet sizes\footnote{We often refer to the culet as the flattened area on the tip of the diamond anvil.} are between 1.2 and 1 mm. As the sample cavity should be a bit smaller than the diameter and rather flat, we have to work with a sample of roughly 500 $\mu$m in diameter and 120 $\mu$m in height, which some NMR spectroscopists might already consider impossible to work with. \\
2) The cavity is tightly enclosed by diamond from two sides, and by a very hard and often metallic gasket which seals the cavity, and provides additional "massive support" of the diamond anvils\cite{Yousuf1982}. Thus, any available free space is located far off the actual sample.\\
3) If we are talking about pressures exceeding 1 GPa, we have to start thinking about hydrostaticity, that is in NMR we need a more or less uniform pressure distribution in the sample cavity, as we are detecting NMR signals from the bulk of the sample. Thus, non-hydrostatic conditions which arise if pressure media turn solid, either at cryogenic temperatures or at high pressures, can lead to ambiguous and distorted NMR spectra.\\
The first NMR experiments in DACs emerged in the late 1980s. The main idea of these pioneering groups was to place a small RF coil operating at predominantly hydrogen frequencies as close as possible to the sample cavity without distorting the diamonds or the metallic gasket. These set-ups include resonators which comprised, for example, a pair of coils placed on the diamonds pavilion\cite{Yarger1995, Lee1989a}, a gradient-field Maxwell coil\cite{Lee1992}, or a single loop cover inductor coupled with a split rhenium gasket\cite{Pravica1998}. A more detailed overview of the development of these high pressure NMR techniques is given elsewhere\cite{Meier2017b}.
\begin{figure}[htb]
  \includegraphics[scale=.12]{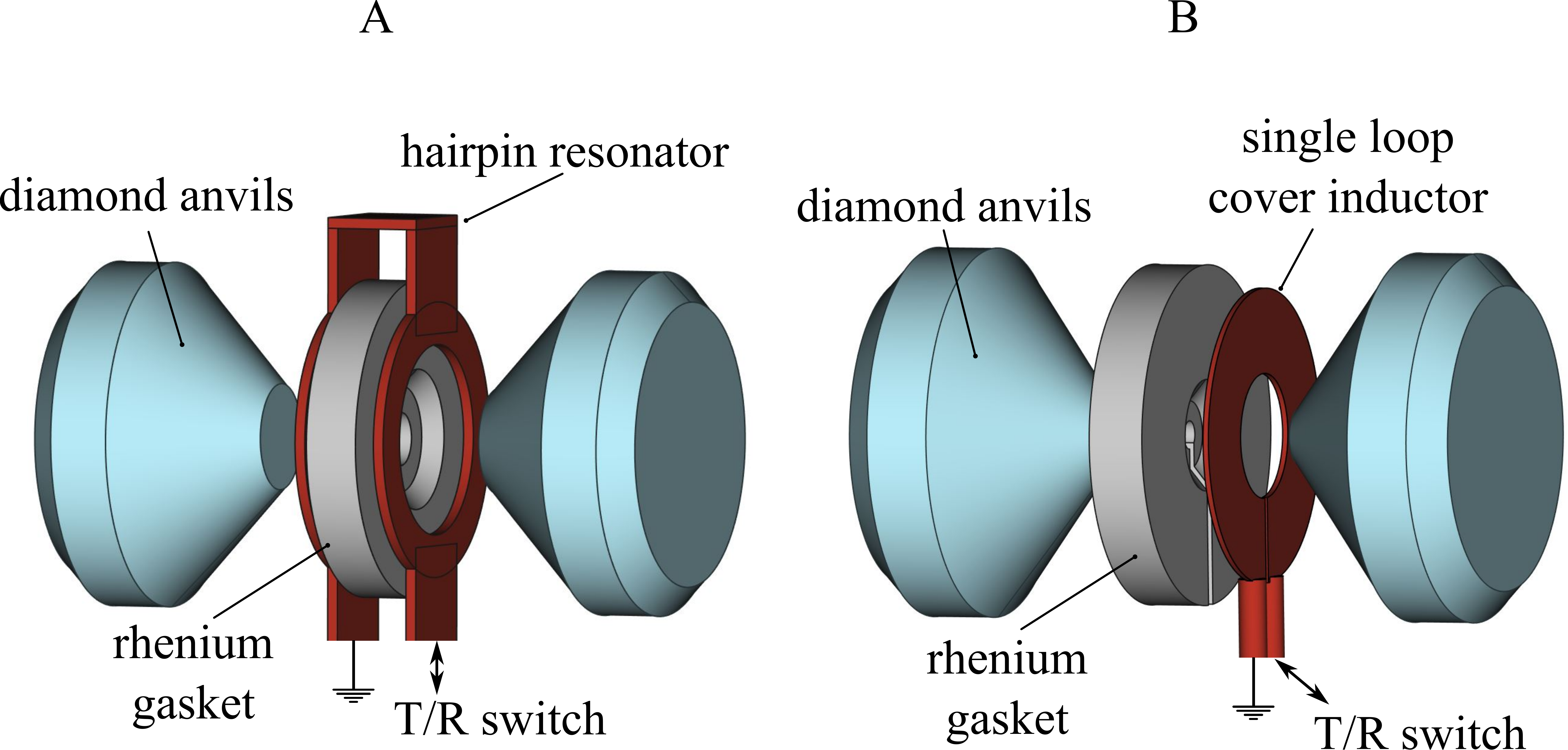}%
 \caption{Two possible arrangements of NMR resonators in a DAC. A) The "Hairpin" resonator could be placed on top and bottom of the rhenium gaskets, thus forming a gradient-field Maxwell coil. After Lee et al.\cite{Lee1992} B) The "Key Hole" gasket resonator basically consists of a copper cover inductor, which is directly connected to the spectrometer and is in electrical contact to a split rhenium gasket, leading to a focusing effect of the RF B$_1$ field at the sample cavity. After Pravica and Silvera \cite{Pravica1998}  
 \label{fig3}
 }
 \end{figure}
 The hairpin and gasket resonator approaches shown in figure \ref{fig3} demonstrate a certain amount of ingenuity needed to overcome the obstacles described above. With these set-ups, pressures as high as 13 GPa could be reached\cite{Pravica1998a}. Nonetheless, the set-ups are far from ideal.\\
As the hairpin resonator is far off the actual samples, the filling factors in this approach are in the order of a fraction of a percent, leading to a spin sensitivity of about $10^{19} \textrm{spin}/\sqrt{\textrm{Hz}}$, which is roughly a factor of 100 lower compared to the standard NMR sensitivity of a static non-DNP experiment. While this improvement sounds very promising, we have to keep in mind that the sample dimensions are already decreased by a factor of about $10^{6}$ in a DAC such as was used in these experiments. Thus, very long data acquisition times hampered a further application of this technique beyond its use for hydrogen NMR in liquid samples, where NMR signals are typically sharp enough to be detected after a couple of thousand scans\cite{Lee1987, Lee1989, Lee2008}.\\
Experience has shown that the problem can only be solved if the RF resonator's filling factor could be significantly improved. At the end of the 1990s, Pravica and Silvera came up with one of the most interesting ideas so far: the electrically conductive rhenium metal gasket was cut open from the sample hole radially outwards, resembling a key hole\footnote{In fact, these resonators are often referred to as "key hole resonators".}. The slit was filled with a mixture of diamond powder and NaCl, which, after careful melting of the NaCl powder, formed a homogeneous filling of the slit. Afterwards, a copper cover inductor, connected directly to the NMR spectrometer, was placed in contact with the slitted gasket. Therewith, NMR experiments could be performed using the electric coupling of both gasket and cover inductor, leading to a locally enhanced B$_1$ in the sample chamber, and an increase of spin sensitivity by one order of magnitude compared to hairpin resonators.\\
However, the downfall of this approach is a bit more subtle. First, as the conductivity of rhenium is one order of magnitude less than that of copper, the quality factor of the key hole resonator gasket is rather low. Coupling of both resonators also turns the copper cover inductor into a lossy resonator. Secondly, the slit in the gasket forms a capacitor with the NaCl grains due to its conductivitiy, breaking down the electric field into small steps of floating potential in the capacitor. Now, since this capacitor  is rather inaccurate, the self resonance of the slit gasket will be far off the desired resonance frequency of the nuclei in the sample, and the slit gasket basically forms a lossy inductance. Furthermore, both sodium and chlorine ions show increasing mobility when under pressure or stress, thus they move due to applied magnetic fields, warming up the gasket through thermal dissipation, and reducing the Q even further. Therefore, all proximity advantages due to the geometry are negated.\\
These developments mark the apex of the first evolution period in DAC-NMR research. Unfortunately, only a handful of groups tackled this demanding task. Also, as sensitivities were very low and actual acquisition times exceedingly long, it was widely believed that this method could only be applied for proton NMR.\\
Nonetheless, while being able to perform NMR experiments at pressures of around 5 GPa seems to be a great accomplishment indeed, we still need to dig deeper into our metaphorical hole in the ground. Far deeper.

\section{Reaching the Lower Mantle: where things become tricky}
\label{micro-coil}

Up to this point, we were merely able to literally scratch the surface of our planet. The analogy of an apple seems fitting: We would have penetrated the apple to just below its peel. So, most of the interesting things are still deeper below, awaiting their discovery.\\
This demand for higher pressures can easily be understood then we think about chemical bonding and crystal structures. To investigate transitions in the electronic or atomic environment of a solid, we need to be able to increase its energy density, i.e. the pressure, up to a point where atomic distances in a system are below a certain threshold, triggering phase transitions. Here, we are typically not only talking about structural phase transitions but also about higher order phase transitions like electronic or magnetic transitions\footnote{which do not necessarily coincide with first order phase transitions.}.\\
Of course, the pressure needed to trigger these transitions changes from system to system. For example, the cuprate high-temperature super-conductors exhibit a layered structure with copper-oxygen layers separating their charge reservoirs\cite{Hazen1987}. These systems are prone to react rather sensitively to the application of pressure. For example, it was reported that the super-conducting transition temperature T$_c$ in   Hg$_{1-x}$Pb$_x$Ba$_2$Ca$_2$Cu$_3$O$_{8+\delta}$ rises significantly under an application of about 30 GPa with a maximum in T$_c$ of 164 K \cite{Gao1994}. Inducing structural phase changes in these systems can potentially occur at pressures far below 10 GPa\footnote{If the experiments were conducted carefully, and special care has been taken to ensure hydrostatic pressure conditions.}, as it was reported in YBa$_2$Cu$_4$O$_8$\cite{Souliou2014}, followed by a complete collapse of super-conductivity\cite{Tissen1991, Mito2014}.
More robust systems like atomic metals, e.g. sodium or lithium, require a much higher energy density before any electronic or structural changes can occur\cite{Ma2009, Han2000}.\\
At this point, DAC-NMR appeared to be stuck in a crisis until into the late 2000s. It was quickly realised that two major problems should be solved, the first being to achieve stable pressures above 10 GPa with good sensitivities, allowing for realistic and time-saving experiments on nuclei other than $^1$H or $^{19}$F, with sample dimensions rapidly decreasing due to the demand for higher pressures. The second issue was a minimisation of the diamond anvils to a point below what was possible at that time in NMR spectroscopy.\\
Apparently, the most promising solution was to use RF micro-coils as close as possible to the sample, even if that would mean to place them directly in the sample chamber. From an NMR perspective, the use of micro-coils is preferable to other methods, as they were shown to exhibit excellent mass sensitivities and large bandwidths due to their small size.\cite{Lacey1999, Olson1995}. Placing such minuscule coils in the pressure chamber of a DAC, however, turned out to be a demanding task. To begin with, the micro-coils would have to be about a factor of 4 to 5 smaller, compared to micro-coils pioneered and characterised before \cite{Stocker1997, Webb1997}. Furthermore, the issue of safely guiding the coil's leads out of the chamber requires either the use of gold liners, which are prone to rupture under stress, or the carving of channels into the metallic gaskets, which is greatly compromising the overall stability of the DAC under load.\\
In 2009 Suzuki et al.\cite{Suzuki2009} presented an intriguing study on the $^{51}$V-NMR of the one dimensional conductor $\beta$-Na$_{0.33}$V$_2$O$_5$ up to 8.8 GPa at cryogenic temperatures. In their figure 1b, a microcoil can be seen placed in the cavity of a Bridgman-type pressure cell\footnote{which has the same working principle as a DAC, but uses metallic anvils often made from non-magnetic WC.}. Unfortunately, the authors did not celebrate this ground-breaking advancement of the field with a separate publication, introducing this approach to a wider NMR community. That was done a short time later in the same year by another group using a strikingly similar set-up\cite{Haase2009}.\\
Both these set-ups were predominantly used by solid-state physicists investigating highly correlated electron systems at low temperatures. In 2011, about the time when I started to work in this field at the University of Leipzig, Meissner et al.\cite{Meissner2011} reported the pressure induced closing of the spin pseudo-gap in YBa$_2$Cu$_4$O$_8$ at pressures up to 6 GPa and temperatures of about 100 K.
\begin{figure}[htb]
  \includegraphics[scale=.2]{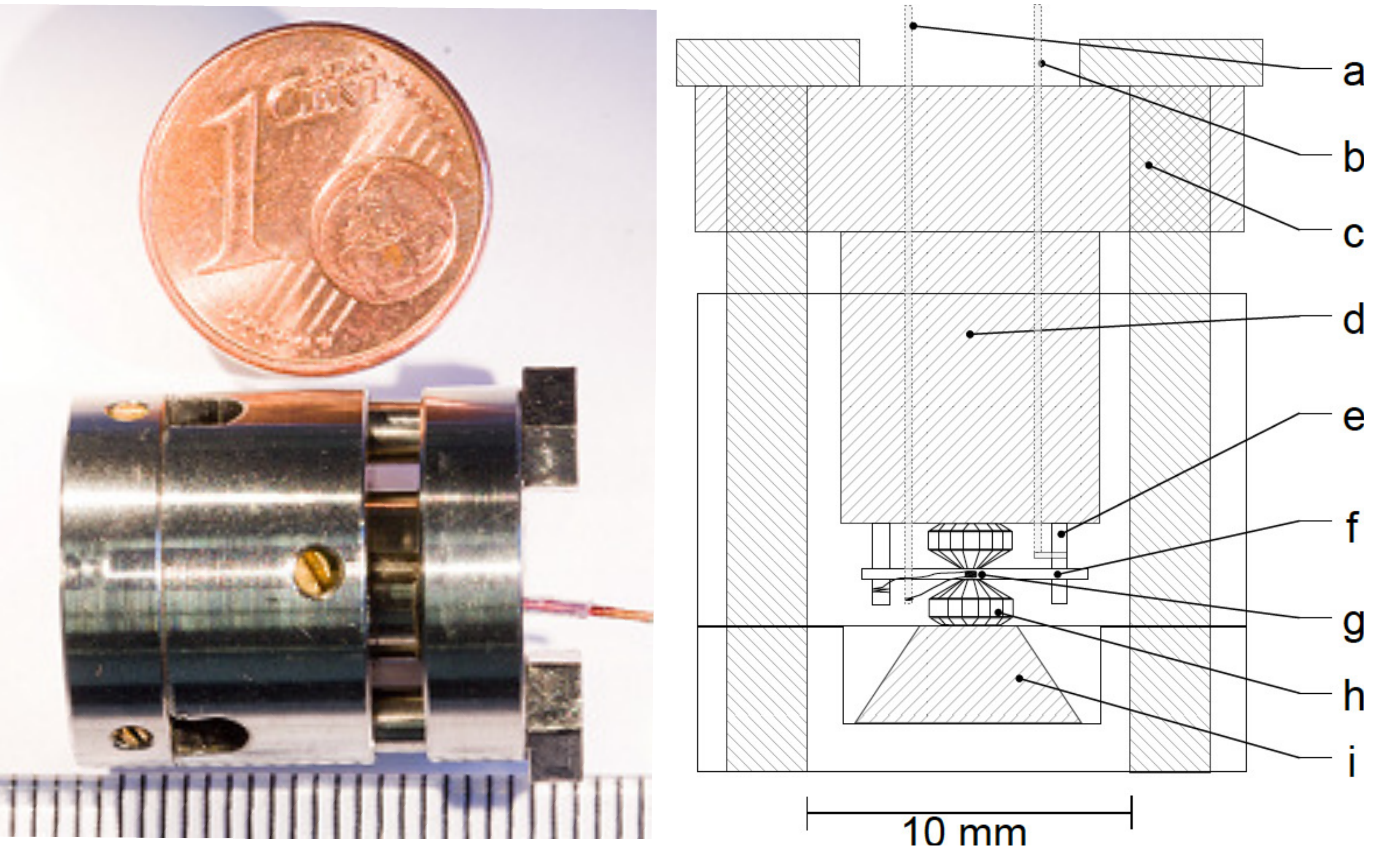}%
 \caption{Left: Miniature non-magnetic DAC made from Titanium-6\%wt Aluminium-4\%wt Vanadium Right: Schematic diagram of all parts of the DAC. Figure from \cite{Meier2014}
 \label{fig4}}
\end{figure}
The pressure cells were manufactured to be fairly small, only 22 mm in length and 18 mm in diameter, and thus could be used in a small bore super-conducting NMR magnet, see figure \ref{fig4}. Similar to the work of Suzuki et al. an average pressure of about 4 to 7 GPa could be realised easily\cite{Meissner2012, Meier2016}.\\ 
Even more important than the achieved pressures, which were comparable to the set-ups discussed in the last chapter,  was the finding that the microcoil set-up yielded very high signal-to-noise ratios, see figure \ref{fig5}, which could be translated to a spin sensitivity of about  $10^{13} \textrm{spin}/\sqrt{\textrm{Hz}}$ which is almost four orders of magnitude lower, and thus much more sensitive compared to the set-ups shown in chapter \ref{midp}\cite{Meier2014a}.\\  
\begin{figure}[htb]
  \includegraphics[scale=.8]{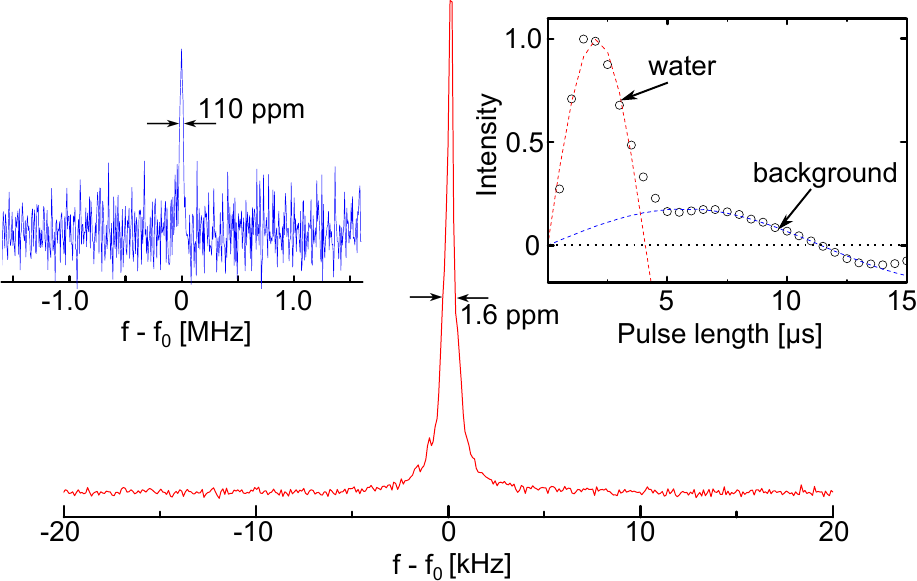}%
 \caption{$^1$H-NMR spectrum after a single shot on water at ambient conditions at a magnetic field of 7 T. Left inset: proton background of the empty cell. Right inset: recorded proton nutation data.
  Figure from \cite{Meier2014}
 \label{fig5}}
\end{figure}
Working in an environment dominated by physicists, we focused our research not on structural determination, or more chemically motivated questions under pressure\footnote{Which will, without a doubt, yield an amazing amount of new phenomena even in the lower pressure range in the near future.}. Our main focus was the change of electronic properties of solids, i.e. pressure induced changes in a solids' band structure, or changes in the occupancy of energy levels of a metals' conduction electrons.\\
If we think about a solid, be it a metal, semiconductor, or insulator, we realise immediately that reducing the inter-atomic distances will inevitably have a large effect on the solids band structure. The most pronounced of these effects is the transition from an insulator to a metal, i.e. a pressure triggered electron delocalisation. Decreasing the distances between atoms often leads to a broadening of valence and conduction bands, culminating in an overlap of both bands triggering electronic conductivity.\\
Of course, such a drastic effect will also significantly influence observable NMR parameters, like spin relaxation, or resonance frequencies.\\
Typically, we can define two distinctively different regimes in NMR. On the one hand, there are insulators, e.g. most organic material is insulating, where the shift of resonance frequency is mainly governed by the diamagnetic shielding of the nuclei by low energy paired-up electrons. Thus, the shift is often small, in the range of some ppm. Spin lattice relaxation is predominantly given by dipole-dipole coupling for $I=\frac{1}{2}$ nuclei or quadrupolar interactions for $I>\frac{1}{2}$ nuclei, and is often in the range from 1 ms up to hours\cite{Bloembergen1948}.\\
On the other hand, as was realised already in the early days in NMR, resonance frequencies in a metal are profoundly higher compared to a non-conducting salt containing the same nucleus\cite{Knight1949}. This so-called Knight shift, named after Walter D. Knight in 1949, is a direct consequence of Pauli-paramagnetism, i.e. the hyperfine interaction of the s-like conduction electrons with the nucleus. This electron-nuclear coupling in fact proved to be so dominant that observable Knight shifts are often two to four orders of magnitude higher than the chemical shifts of the insulating compounds of a given metal. Furthermore, Korringa found\cite{Korringa1950}, based on nuclear relaxation theory from Heitler and Teller\cite{Heitler1936}, that the spin lattice relaxation times in a metal must also be directly correlated to the hyperfine interaction felt by the nucleus. The famous Korringa relation combines both Knight shift $K$ and T$_1$,  and shows that the ratio of  $K^2$ and T$_1$ at constant temperatures only depend on natural constants, and the gyromagnetic ratios of the electron and the nuclei. As the Korringa relation should also be independent of volume, it is a perfect tool to probe and identify metallisation processes.\\
Coincidentally, we were given a sample of nano-crystalline AgInTe$_2$ powder at the time, synthesized by our chemistry department in Leipzig University \cite{Schroder2013, Welzmiller2014}. This compound, which is semiconducting at ambient conditions, was believed to become fully conducting at the chalcopyrite to rocksalt structural transition\cite{Range1971}, occurring in a pressure range between 4 and 6 GPa.\\
First experiments on AgInTe$_2$ powdered samples, which has not been characterised by NMR so far, showed that the $^{115}$In spectra displayed a first order quadrupole interaction \--- indium is nuclear spin 9/2 \--- with a quadrupole frequency $\nu_q$ of about 45 kHz, thus the 8 satellite transitions were found to be heavily broadened and merged into a broad symmetric background around the sharp central transition, see figure \ref{AgInTe}. The spectra were found to be relatively strong shielded, having chemical shifts of about -400 ppm relative to an aqueous solution of an indium salt. Furthermore, T$_1$ relaxation times were found to be in the range of some 10 ms, indicating relaxtion mechanisms governed by quadrupole interaction.\\
\begin{figure}[htb]
  \includegraphics[scale=.7]{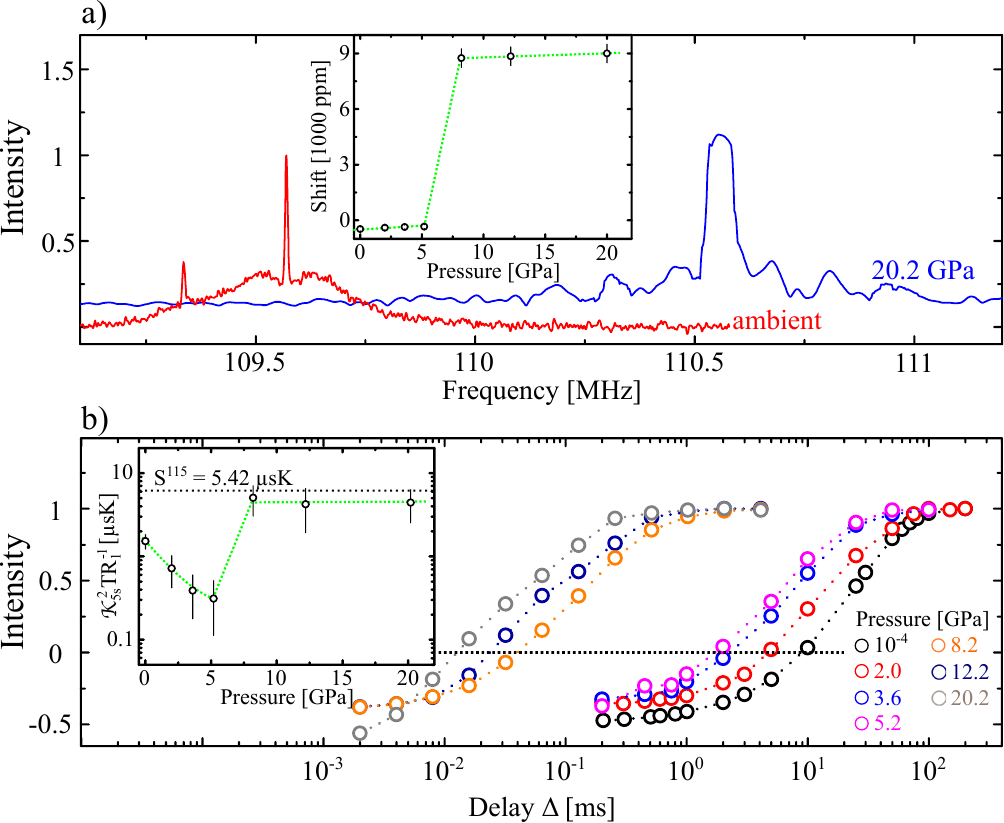}%
 \caption{Top panel: Recorded $^{115}$In-NMR powder spectra recorded using a quadrupole echo sequence at ambient pressure (red) and at 20 GPa (blue). The vertical shift was introduced for better comparison. Inset: Obtained frequency shift values recorded over the full pressure range in these experiments. Bottom panel: Magnetisation recovery curves obtained during inversion recovery experiments as a function of the separation pulse $\Delta$ between the 180$^{\circ}$ inversion pulse and the detection pulses. Inset: Korringa ratio as a function of pressure. The dotted line depicts the expected values from a free electron metal of Indium atoms. 
 \label{AgInTe}}
 \end{figure}
 Up to about a pressure of 4 to 5 GPa, these parameters were not found to change significantly. Above 5 GPa, however, both the resonance shift as well as T$_1$ changed rather drastically by about 9000 ppm higher in frequency, and two orders of magnitude faster relaxation times. Combining both effects, the Korringa relation was found to suddenly become volume independent, and not change in a pressure range from 8 to 20 GPa, indicationg electron delocalisation in AgInTe$_2$.\\
 Up to this point, we had not payed much attention to the problem of hydrostaticity for NMR experiments at pressures above 7 GPa. At these compressions, most of the commonly used pressure transmitting media, like glycerol or Daphne, are solidified. This leads to so-called dry contact of the diamond faces with the gasket and sample, and will result in pronounced pressure gradients. Unfortunately, the influence of non-hydrostatic pressure conditions on NMR spectra or relaxation mechanisms has not been investigated in detail so far. Nevertheless, there is mounting evidence of the importance of this issue for some systems.\\
 A good illustrative example is the behaviour of metallic aluminium under pressures up to 10 GPa. In 2014, Meissner et al.\cite{Meissner2014} presented experiments on the $^{27}$Al-NMR spectra of metallic Al powder. There, the authors claimed that the observed deviation of the Knight shift and the sudden increase in linewidth at about 4 GPa must be due to a so-called Lifshitz transition, which occurs if a van-Hove singularity of a given energy band in the solid's band structure crosses the Fermi energy E$_F$, and becomes partly filled or unfilled. One might argue that such transitions should be ubiquitous in solids under pressure, as the band structure typically changes quite significantly under compression. However, direct experimental observation of such an effect has been scarce, because these effects are typically smeared out by thermal excitations close to E$_F$. Thus, they should only be observable at low temperatures of about 10 K or below. Unfortunately, no low temperature experiments could be published confirming these findings at 300 K.\\
 Careful re-examination of the experimental conditions, however, led to a slightly different, and much more simple, interpretation of these findings. In fact, the 'smoking-gun' evidence correlating the transitions found with the experiment itself was that 4 GPa, the pressure where both observed effects on the $^{27}$Al spectra became dominant, coincides with the reported crystallisation point of the glycerol pressure medium used. Thus, new sets of experiments using paraffin oil as a pressure medium, which solidifies at much higher pressures of about 12 to 13 GPa, showed that both the deviation of the Knight shift as well as the increase in linewidth strongly followed the onset of non-hydrostatic pressure conditions. A more detailed account for these effects are given in \cite{Meier2017b}.
 Thus, by now, every NMR spectroscopist should be aware of the deceptional effects occuring when at sufficiently high pressures, their pressure media begin to cristallise.\\
 \begin{figure}[htb]
  \includegraphics[scale=.8]{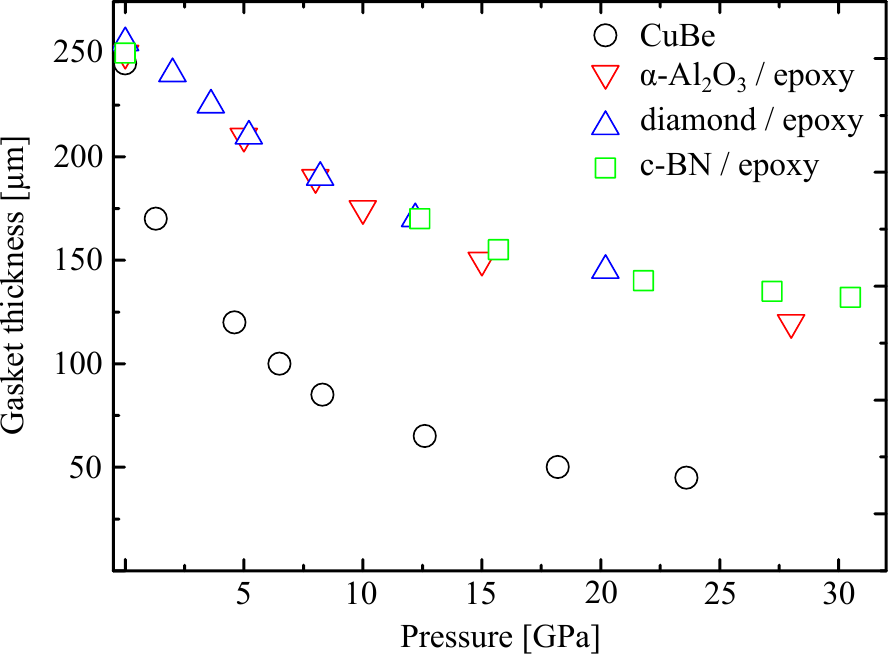}%
 \caption{Height of the sample cavity for different pressures for various gasket materials. Taken from \cite{Meier2015a} 
 \label{heights}}
 \end{figure} 
 Within the five odd years of my working with micro-coils, some serious limitations became obvious. Due to the limited space available in the sample chambers in a DAC, very thin insulated wires had to be used to prepare the coils. Of course, companies providing thin insulation wires made from copper or gold are sparse, and acquiring larger amounts often rather expensive. Furthermore, only organic insulating materials were possible to deposit on the wires, thus limiting their application to low $\gamma_n$ NMR nuclei, because hydrogen backgrounds excessively overlapped with $^1$H-NMR spectra. Finally, copper wires sold by \textit{Polyfil} could be acquired. With these 18 $\mu$m thick wires, insulated with Polyurethane, coils of 3 to 6 windings could be manufactured. Thus, the coils had approximate dimensions of 200 \-- 500 $\mu$m in diameter and 80 to 160 $\mu$m in height. To reach higher pressures, smaller culet faces must be applied, which reduces the initial sample cavity quite significantly. This leads to certain boundaries of the applicability of micro-coils in DACs. To give one example, using a pair of two 500 $\mu$m culeted anvils, reaching about 40 GPa on average, requires a sample volume of 160 $\mu$m in diameter and 40 to \-- 50 $\mu$m in height for best stability; but this would require micro-coils to be made having only two turns, which is almost impossible to manufacture.\\
 \begin{figure}[htb]
  \includegraphics[scale=.3]{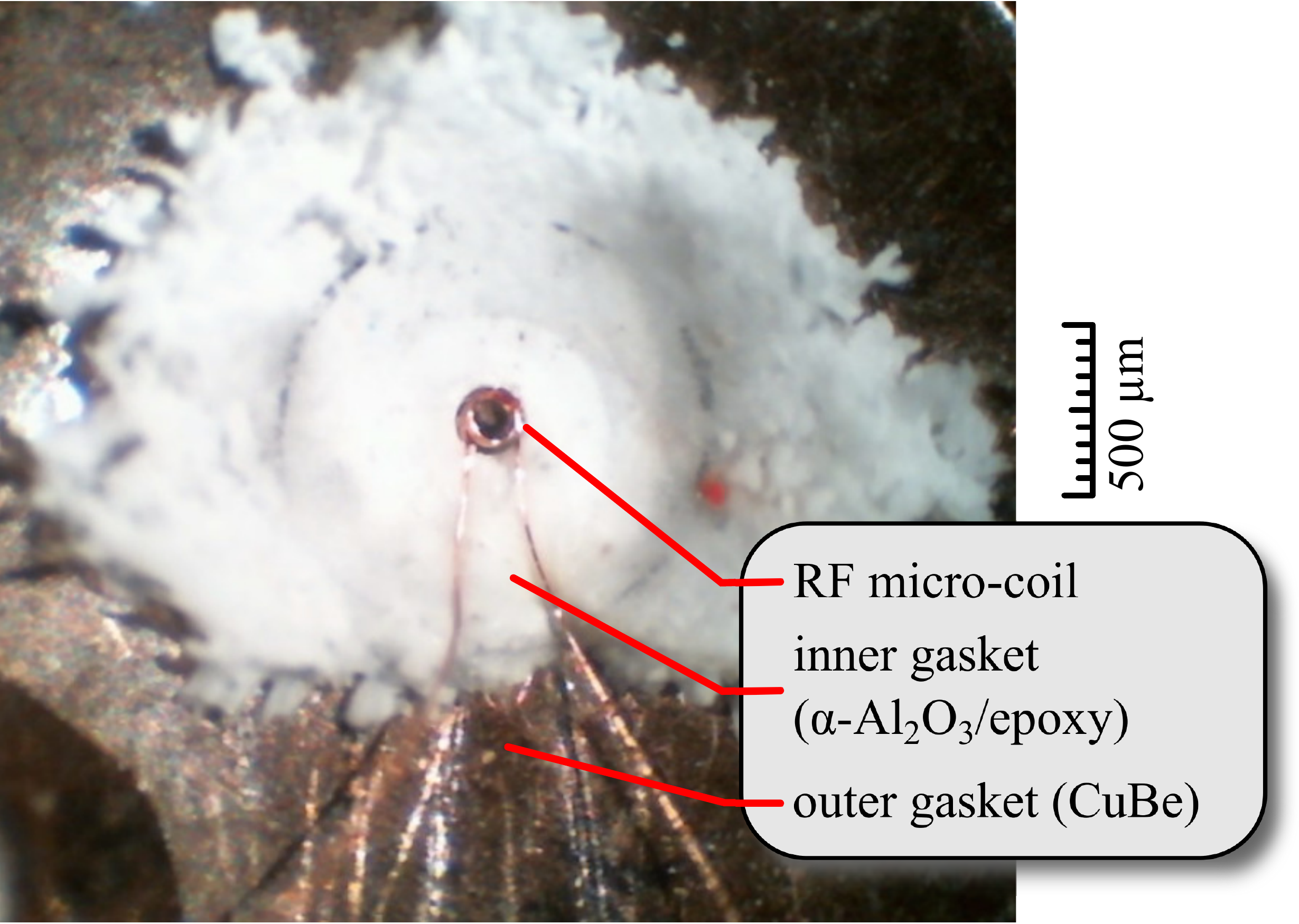}%
 \caption{Photograph of the composite gasket assembly. The diamonds used in this photograph were culeted to 600 $\mu$m. Taken from \cite{Meier2015a}
 \label{gasket}}
 \end{figure} 
 Another serious problem originated in the use of the gasket material. Due to its low magnetic susceptibility, Cu-Be chips were used as gaskets blanks. Unfortunately, this alloy turned out to be quite soft under compression, leading to sample height reductions of almost a factor of four within some GPa, see figure \ref{heights}. Such a pronounced cavity collapse would lead to significant deformations of the RF micro-coils, leading to B$_1$ field inhomogeneities and, thus, reduced sensitivity within a single pressure run.\\
Therewith, the quest for a method for gasket stabilisation was on. A possible solution was found by using so-called composite gaskets, which are made by replacing the pre-indented part of the gasket with a rigid matrix of a nano-crystalline ultra hard material, like diamond or c-BN. A photograph of such a gasket design is shown in figure \ref{gasket}. Here, an amorphous mixture of $\alpha-Al_2O_3$ and epoxy was used within a DAC, using a pair of 600 $\mu$m anvils.\\  
It could be shown that these gaskets allow for significantly stabilised cavities as illustrated by the much bigger recovered gasket heights in figure \ref{heights}. Figure \ref{FID} shows time domain solid echoes, as well as their Fourier transform, at pressures up to 30 GPa. 
\begin{figure}[htb]
  \includegraphics[scale=1.3]{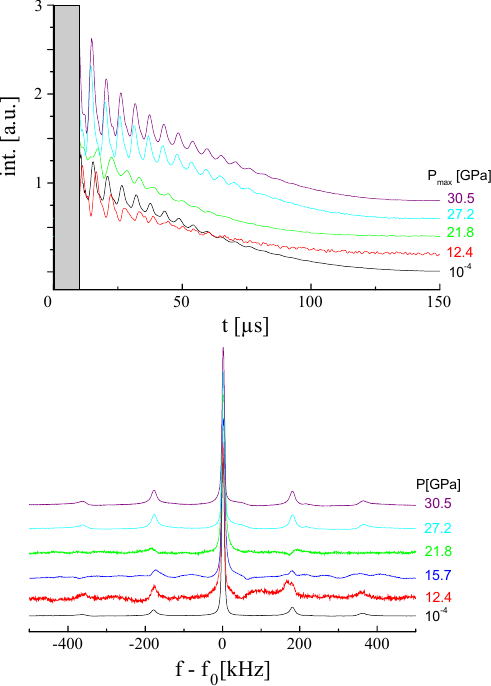}%
 \caption{Upper part: $^{27}$Al solid echoes of Cr:Al$_2$O$_3$ up to 40 GPa. The spectrometer was blanked off for 5 $\mu$s. Lower panel: Corresponding NMR spectra after Fourier transform. Both spectra and time domain signals were offset in the y axis for better comparison. 
 \label{FID}}
 \end{figure} 
As can be seen, even at 30 GPa, features in both time and frequency domain remain sharp, while at the same time allowing for rather high excitation bandwidths.\\
These experiments mark the high point in micro-coil high pressure research. The difficulty in preparing the set-ups, however, greatly limited the applicability of this approach, since they seem to resemble an art form rather than a reliable and reproducible method of science. Also, the results shown in figure \ref{AgInTe} and \ref{FID} can actually be considered singular events, and an average pressure limit of about 7 to 8 GPa using micro-coils in DACs would be reasonable. Unfortunately, most prominent questions in contemporary physics, chemistry and the geosciences appear to happen at considerably higher pressures. Thus, in order to reach even deeper into the Earth, completely new resonator structures needed to be developed.\\
The Mega-Bar regime awaits!

\section{Think Mega-bars!}
\label{mbar}

\begin{figure}[htb]
  \includegraphics[scale=.17]{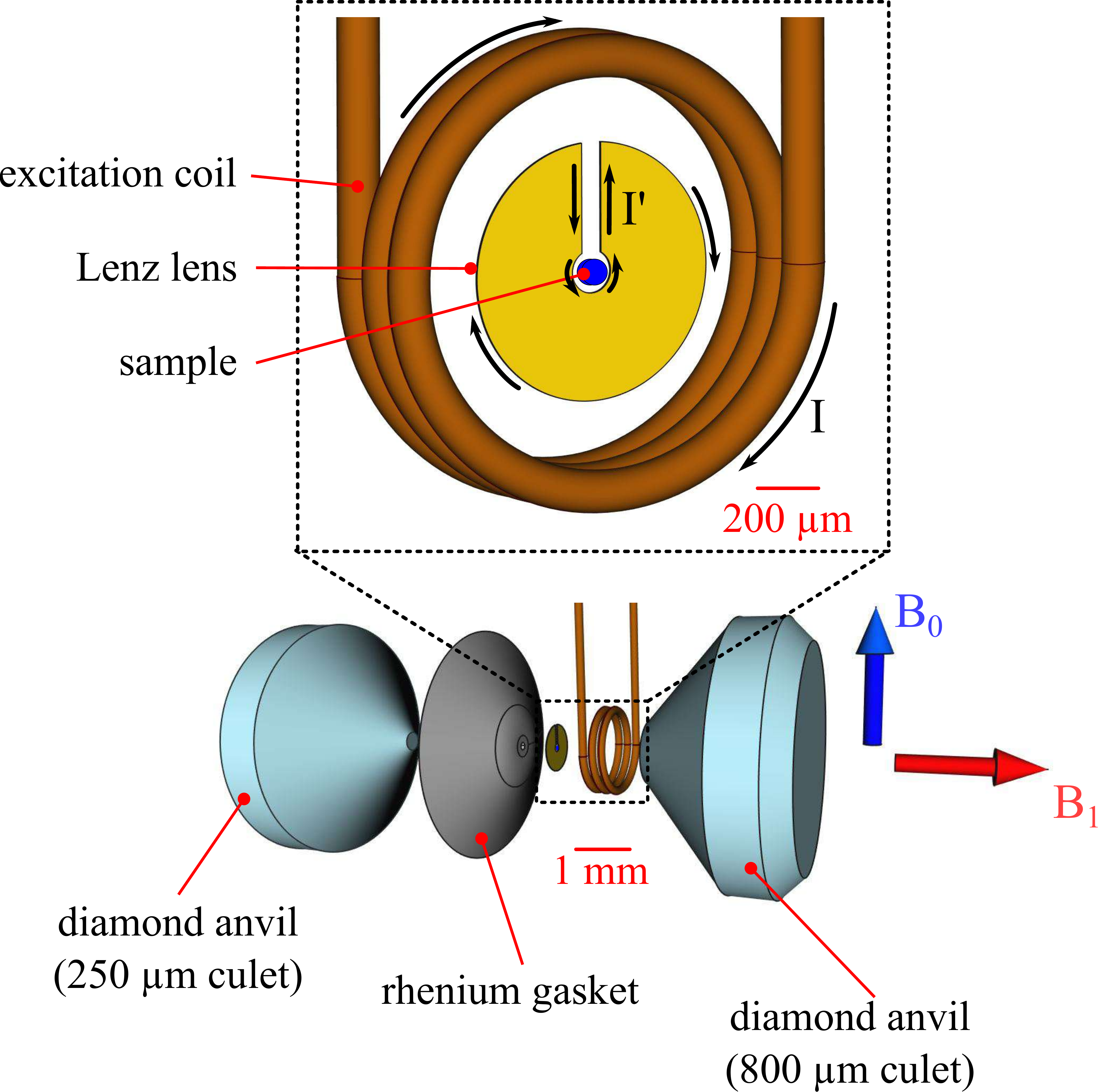}%
 \caption{Schematic explosion diagram of the resonator setup and the anvil/gasket arrangement. The blue and red arrows denote the directions of the external magnetic field B$_0$ and the RF magnetic field B$_1$, respectively, generated by the excitation coil and the lens, which is compressed between the rhenium gasket and the 800 $\mu$m culeted diamond anvil. The enlarged picture shows the RF arrangement of the excitation coil with the Lenz lens. Black arrows denote the directions of the high-frequency current.
Taken from \cite{Meier2017}
 \label{SciAdvfig1}}
 \end{figure}

\begin{figure*}[htb]
  \includegraphics[scale=.25]{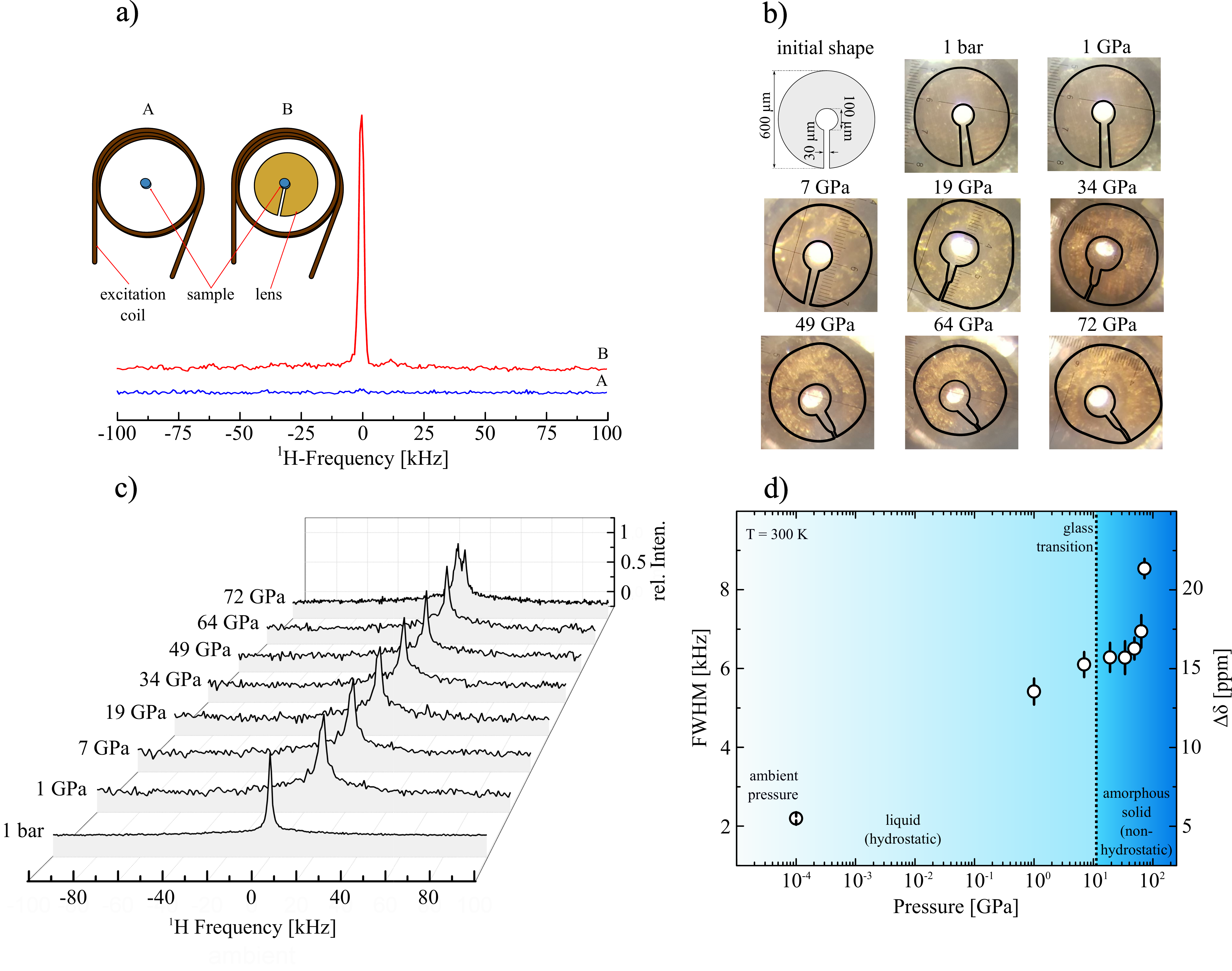}%
 \caption{Sensitivity and stability tests of the Lenz lens resonator setup in a DAC. (A) Proton spectra of paraffin at ambient pressure with and without the use of a Lenz lens. (B) Photographs of different deformation states of the lens under pressure. Black contours are guide to the eye. (C) Recorded $^1$H NMR spectra. At ambient conditions, 100 scans were accumulated, whereas at higher pressures, only single-shot spectra after a single $\pi/2$ pulse were recorded. (D) Pressure dependence of the FWHM linewidths. The dotted line denotes the crystallisation pressure at ambient temperature, and the shaded areas denote the liquid and amorphous phases of paraffin. The glass transition pressure was obtained from other methods
Taken from \cite{Meier2017}
 \label{SciAdvfig567}}
 \end{figure*}

\begin{figure*}[htb]
  \includegraphics[scale=.26]{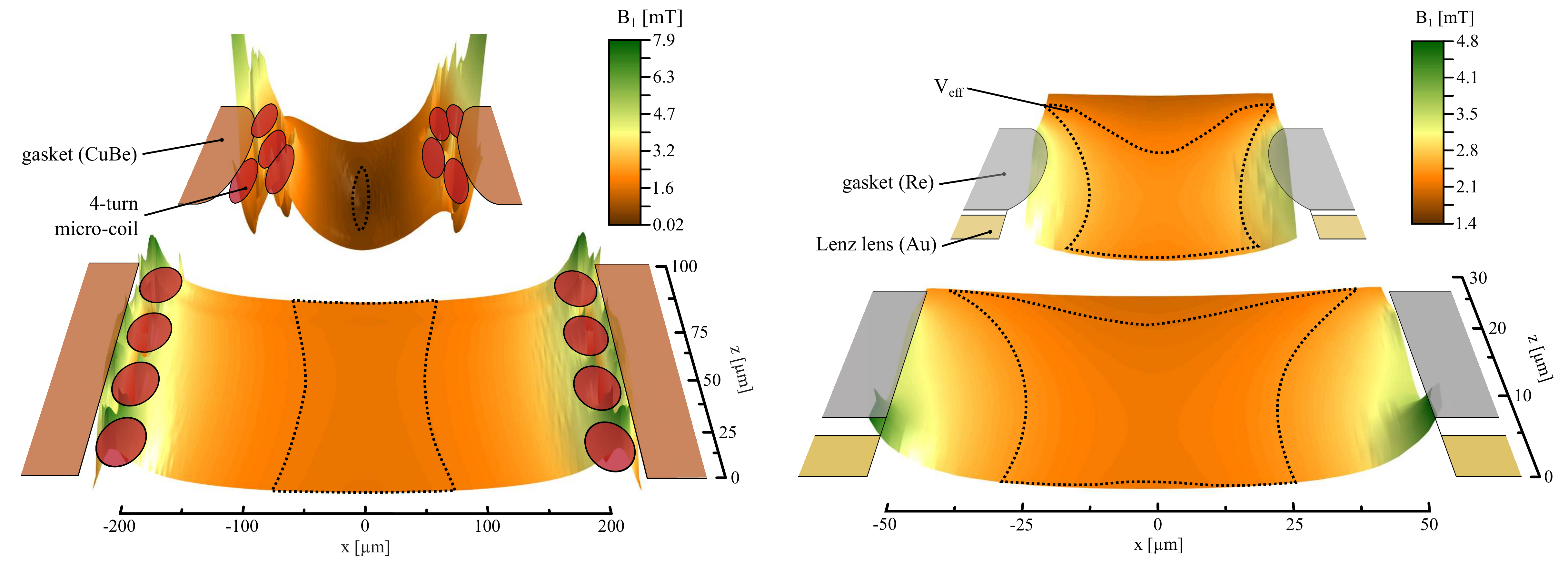}%
 \caption{Magnetic field maps of the B1 fields generated by a micro-coil of four turns (400 $\mu$m in diameter and 100 $\mu$m in height, left), and of a Lenz lens made from a solid sheet of gold foil (right). Above the initial configurations, the figure shows two deformed set-ups at a stage where the initial height and diameter of the sample cavity are reduced by 50\%, occuring well below 10 GPa when bigger CuBe gaskets were used. Using flat rhenium gaskets, this state of deformation typically occurs at substantially higher pressures. Indicated as an overlay are the geometries of both RF resonators and the corresponding gaskets. The deformed state of the micro-coil was reproduced from a photograph of an opened pressure cell working up to 6 GPa. Different gasket materials were used in both set-ups, which were also included in the simulations. The dotted
lines represent the effective observable sample volume.
Taken from \cite{Meier2017}
 \label{fig9}}
 \end{figure*}

\begin{figure}[htb]
  \includegraphics[scale=.17]{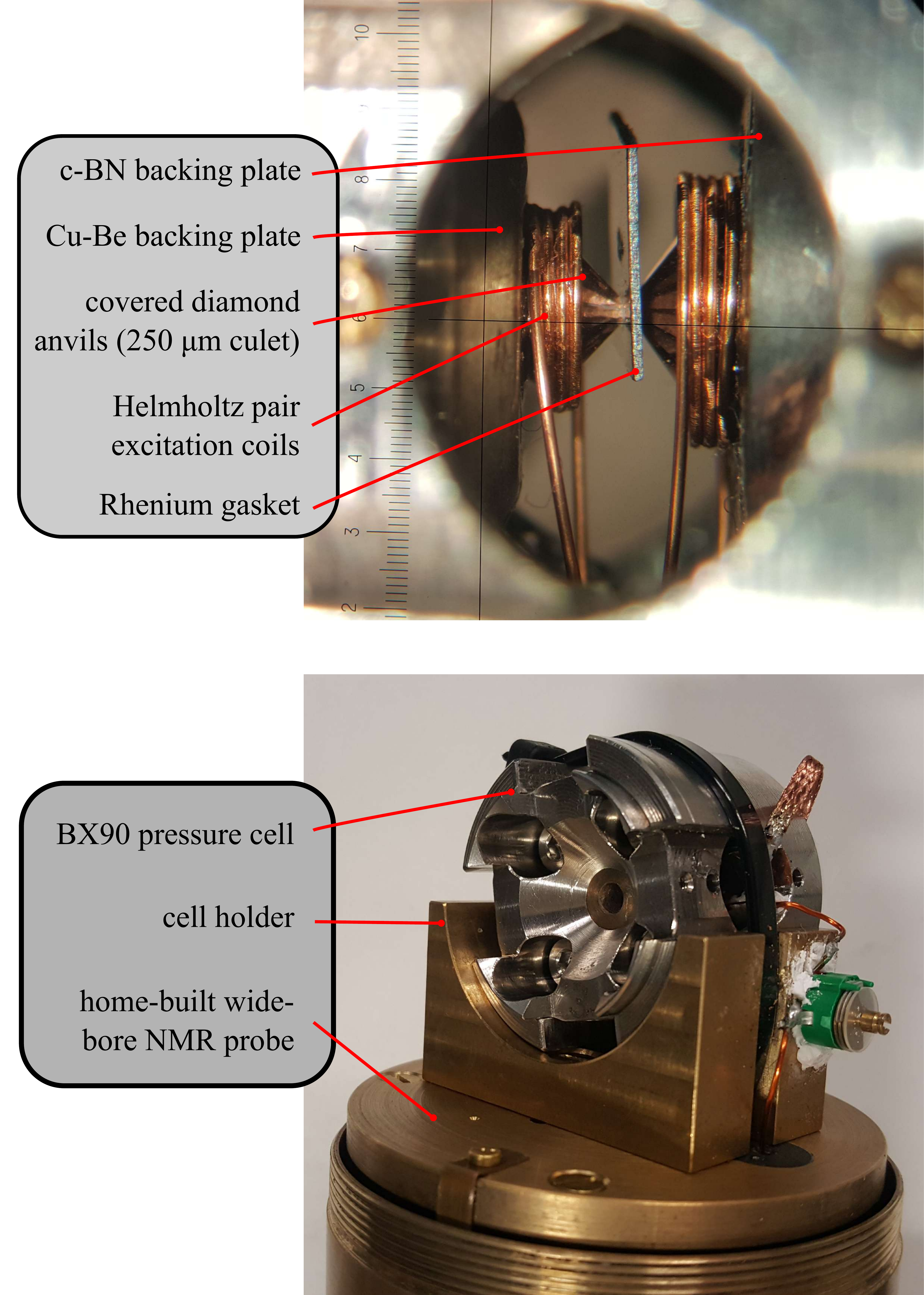}%
 \caption{Upper photo: completely prepared DSLL-resonator for a DAC equipped with two 250 $\mu$m culeted anvils. Lower photo: mounted pressure cell on a home-built NMR probe
 \label{cell}}
 \end{figure}

\begin{figure}[htb]
  \includegraphics[scale=.3]{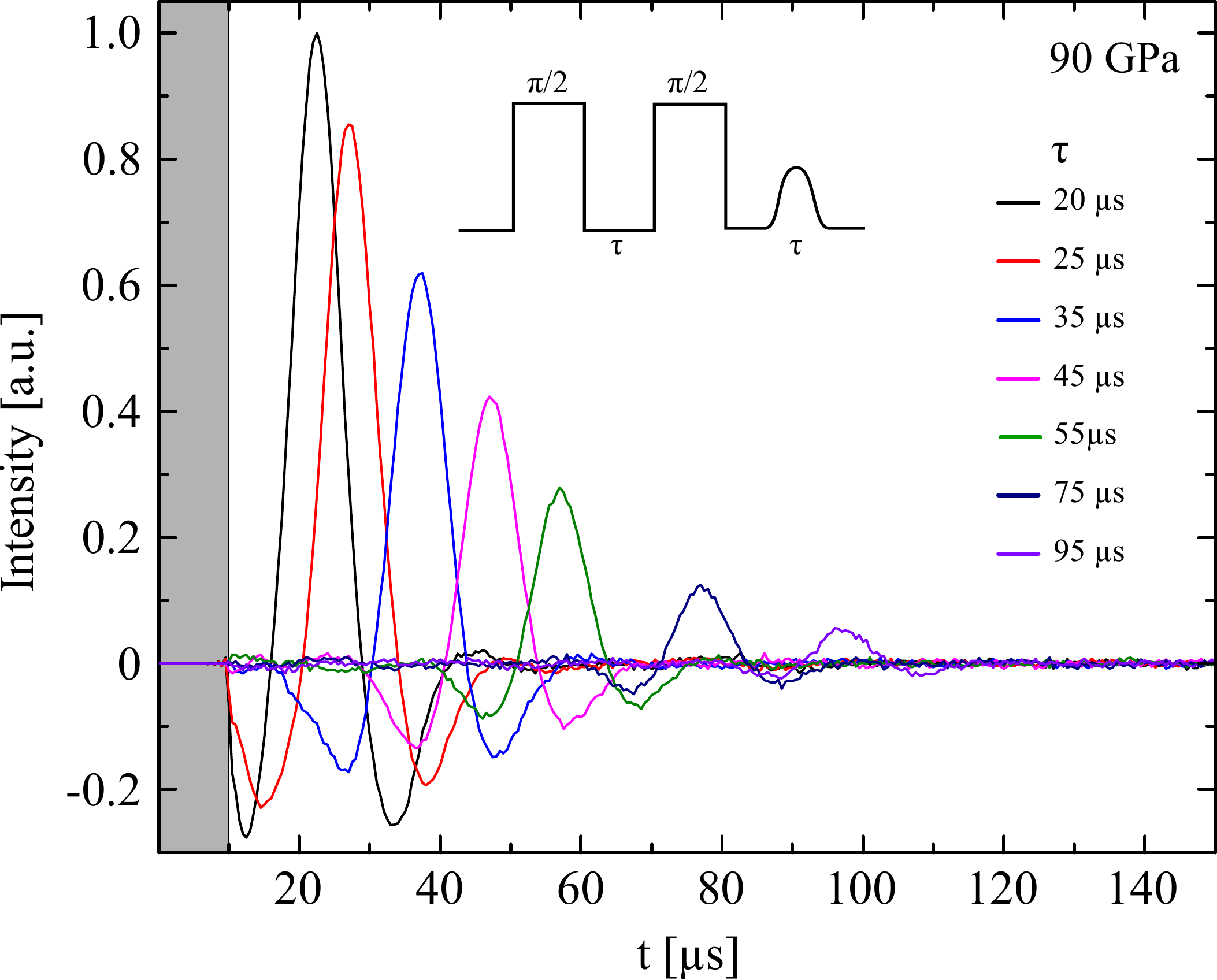}%
 \caption{Recorded solid echo train of high pressure ice X at 90 GPa. The spectrometer was blanked off for 4.6 µs after the second 90° pulse (grey area). Taken from \cite{Meier2018a}.
 \label{echo}}
 \end{figure}

Our journey is almost at an end. Despite what the title of this chapter suggests, NMR experiments above 1 Mbar, or 100 GPa, have not been realised so far. But it is a close call.\\
In the course of the last year, high pressure NMR gained momentum, and routinely reaching pressures above 30 to 40 GPa could be feasible even for a broader NMR community. But let us start at the beginning.\\
At the end of the last chapter I have summarised attempts to conduct magnetic resonance at pressures well above 10 GPa. This research also coincided with me obtaining my PhD, and moving on to a purely high pressure oriented institute. Here, it instantly became clear that pressures well above the state-of-the-art must be realised. Having gathered all the experience from implementing micro-coils in DACs, we began looking for a completely different approach.\\
Experimenting with planar micro-coils or even micro-striplines, it quickly became apparent that something much more robust must be used, as both micro coils and striplines did not sustain the exceedingly high deviatoric stresses in a DAC made for Mbar measurements. The revelation came in form of magnetic flux tailoring Lenz lenses (LL), pioneered by the group of Prof. Korvink in the KIT\cite{Spengler2017, Korvink2017}.\\
These passive electro-magnetic devices are governed by Lenz' law of induction, and could be shown to locally amplify the RF B$_1$ field at the sample chamber. This very handy focussing effect is predominantly given by the geometry of these lenses. In a two dimensional plane, an outer winding builds up current, which is induced within a small region along the rim of the lens by an excitation coil. The current is fed into the inner part of the lens, where an anti-winding deposits the magnetic field within the LL's inner diameter\cite{Jouda2017}. Any sample within this inner hole would consequently feel a much higher B$_1$ than without a lens. In this sense, these LLs work as a flux transformer, and display an astonishing degree of flexibility in terms of their field of application.\\
Of course, we immediately tried to implement these fascinating devices into one of our pressure cells. We used not a symmetric arrangement of diamond anvils at first, but two anvils of much different culet sizes. These so-called indenter cells are often used for feeding small wires into the sample cavities, as the gasket only deforms in the direction of the sharper anvil, leaving a lot of space under the gasket open for further manipulations. Thus, we decided to use a 800 $\mu$m base and a 250 $\mu$m primary anvil. The LLs were cut out of 5 $\mu$m thin gold foil \--- 600 $\mu$m outer diameter and 100 $\mu$m inner diameter \--- carefully placed at the centre of the base anvil, and aligned with the 100 $\mu$m sample hole of a rhenium gasket. The gasket was covered with a thin layer of korundum in order to electrically insulate it from the LL. The excitation coil was placed directly on the pavilion of the base anvil to achieve sufficient inductive coupling into the LL. Figure \ref{SciAdvfig1} shows this set-up.
This set-up proved not only to be exceedingly stable under compression, but also to yield much higher sensitivities compared to all other methods so far. Figure \ref{SciAdvfig567}a shows one of the first test scans using paraffin oil as a sample. As can be seen, in the case when no LL is used (i.e. NMR experiments only performed with the bigger excitation coil residing at the base anvils pavilion), SNR is very poor, in the order of $10^{-3}$ after a single scan. Using the LL, however, leads to a significant increase in SNR by almost four orders of magnitude, corresponding to detection limits of only $10^{12}$ spins/$\sqrt{Hz}$. 
The lenses were found to be mechanically stable under compression, see \ref{SciAdvfig567}b. The overall shape of the LL was kept more or less intact up to a pressure of 72 GPa. Further increase in pressure led to the destruction of the anvils. In \ref{SciAdvfig567}c, $^1$H-NMR spectra of paraffin at increasing pressures of up to 72 GPa are shown. These spectra are all acquired after a single scan and demonstrate impressively how the spin sensitivity basically remains constant under compression; this is in drastic contrast to what was observed with the micro-coil set-up.\\
The attentive reader might have recognised that the spin sensitivities realised with the LL resonators are about 2 up to 4 orders of magnitude lower, thus more sensitive, compared to the micro-coil set-ups. The reason for this might strike some as trivial, but I would like to explain it nonetheless. Consider using a micro-coil  of 150\---200 $\mu$m in height and about 300\--400 $\mu$m in diameter in a DAC equipped with two 800 $\mu$m anvils. The sample cavity in such a DAC is most often much bigger than what is considered a 'safe' pressure cell\footnote{In fact, we often prepared DACs like this to use as much sample as possible due to limited sensitivity.}.\\
Under compression, of course, such a DAC can be at best considered rather unstable, and the sample cavity is prone to collapse due to significant flow of the Cu-Be metal. Thus, every micro-coil will be subject to immense deformations, leading to a decrease in the 'effective' sample cavity. To underline this thought, numerical simulations have been conducted of the RF B$_1$ fields in a micro coil set-up as well as for a LL-resonator, see figure \ref{fig9}. \\
In accordance with similar calculations from van Bentum et al.\cite{VanBentum2007}, the B$_1$ field map in the x-z plane of a flat micro-coil, with a length-to-diameter ratio of less than unity, the magnetic field is fairly inhomogeneously distributed, with the highest magnetic fields close to the respective windings. The effective observable sample volume, V$_{eff}$, with a B$_1$ homogeneity within 20\% of the central field, is about 1.7 nl, which is about 14\% of the total available sample space, and stores only 6\% of the total magnetic field energy of the micro-coil. Moreover, as indicated by the “deformed resonator”, the B$_1$ field homogeneity greatly suffers from an irregular arrangement of the current-carrying wire segments of the micro-coil. This deformed state typically arises already at relatively low pressures. Depending on the choice of gasket materials and the geometry of the sample cavity, a collapse can occur at relatively low pressures of some GPa, which will lead to significant deformations of the interior of the cavity, including the placed micro-coils. In this particular case, V$_{eff}$ drops to a 1/20th of a percent because of significant B$_1$ field inhomogeneities, while at the same time storing only about 0.003\% of the total magnetic field energy. In addition, at these compressions, the risk for coil gasket or inter turn short circuits increases rapidly\cite{Meier2016}, rendering the application of micro-coils in DACs increasingly unreliable above 10 GPa.\\
In the case of the Lenz lenses, on the other hand, the RF B$_1$ field
appears to be homogeneous over more than at least 40 to 50\% of the total sample cavity, storing about 30\% of the magnetic field energy. Strikingly, under compression, the situation does not deteriorate significantly, and both the stored energy (≈35 \%) and V$_{eff}$(≈ 47 \%) remain almost constant. The B$_1$ field strengths of the LL resonators found in the simulations compare well with the actual field strengths found via nutation experiments, which is further evidence of the applicability of this approach.\\
Application of LLs in DACs proved to be a real game changer. From now on, diamond anvil cells of standard design could be used, closely following well established preparation guide lines\cite{Eremets1996}. This opens, at least in principle, the door for NMR at 100 GPa!\\
Let us proceed to the most recent technical advancement. Using a diamond indenter cell, as in the aforementioned case, poses some limitation on the maximal reachable pressure ranges (if two diamonds of different culets typically have a somewhat lower maximal pressure than two identical diamonds). For example, two diamonds of 500 $\mu$m culets might reach pressures of about 40 GPa. But if we were to substitute one anvil with a 700 $\mu$m culeted anvil, the maximal pressure could drop to below 30 GPa. Thus, if we want to penetrate the Mbar regime, we need to develop a method for using the LLs in a symmetrical arrangement of anvils.\\
With the introduction of the so-called Double Stage Lenz Lens (DSLL) resonator, this problem could be solved\cite{Meier2018a}. The set-up uses a thin layer of deposited copper on the complete surface of the anvils. A structure of two lenses was cut into this layer using a focused ion beam. The first stage LL is situated along the pavilion of the anvil, with its inner diameter slightly below the diamond's culet face. The second stage LL is the main driving LL and lies directly on the culet. It has an outer diameter of 250 $\mu$m and an inner diameter of 80 $\mu$m, closely following the sample hole diameter. The DSLL resonator is driven by a multi-turn copper coil of 4 mm in diameter placed around the diamond anvil. 
This resonator structure was also realised on the other anvil. After loading and pressurising, both driving coils were connected to form a Helmholtz coil arrangement. Figure \ref{cell} shows photographs into the pressure cell equipped with such a DSLL resonator, and a BX90 pressure cell on a wide-bore NMR probe. \\
The working principle of these DSLL-resonators is similar to the single LL resonators with an additional LL to further focus the B$_1$ field at the sample chamber. One could ask why an application of a single LL would not work as well; the reason is that a single LL running over the sharp edge of the culet would have been cut off under compressional strain. Furthermore, as this ripping off would most likely not occur in an evenly matter, the complete resonator would inevitably turn into a lossy inductor.
First experiments have been performed recently on an initial sample of 100 pl of water. Figure \ref{echo} shows an obtained solid echo train at 90 GPa. For these experiments, only 1000 scans were accumulated, resulting in a SNR per shot of about 39, and a time domain limit of detection of $5\cdot10^{11}$spins/$\sqrt{Hz}$, which is about a factor of two lower than in the case of a single LL.
This last development is particularly easy to prepare, as no time consuming alignment processes, or meddling with micro-coils, is necessary. Furthermore, these are the first experiments with a standard DAC allowing to access the complete available pressure range of the chosen diamond. In this case, 90 GPa is widely considered to be the limit for standard Drukker-type diamonds of 250 $\mu$m without a bevel. In fact, X-ray absorption measurements on these cells at 90 GPa have shown the diamonds to be heavily stressed, leading to a cupping of the culet faces to a degree that both diamond rims are almost touching one another. Thus, a further increase in pressure will certainly result in complete destruction of the anvils.\\
As already stated at the beginning of this chapter, the border to the Mbar regime has not been passed yet, but the recent developments leaves one to argue that this might only be a matter of time. Of course, application of smaller culet sizes, and thus higher pressures, will inevitably reduce the available initial sample volume; but on the other hand, the sensitivity of the DSLL-resonator will benefit from the increased proximity due to smaller inner diameters of the second stage LL, as well as a reduced distance between both pairs of LL on the advancing diamonds.\\
\section{A Few Last Words}
\label{end}
Within this admittedly rather short overview about the technical evolution of high pressure NMR devices, we came across an extraordinary degree of ingenuity. Every little step towards the final installments has certainly been difficult, riddled with mishap and frustration. Nonetheless, the pressure possible nowadays surpasses what was thought possible some years ago almost a hundred fold.\\
It is often said that progress in diamond anvil cell research is by no means revolutionary but evolutionary, every advancement in the field comes together with a new problem longing for a solution. This thought seems to be particularly accurate for high pressure NMR. \\
At this point, DAC-NMR advanced both in terms of NMR sensitivity and in achievable pressures to an amount which was widely believed impossible. In fact, over the course of my relatively short academic career, I was met with both disbelieve and ridicule about my vision to perform NMR above 100 GPa. But, as years went by and the field went forward, the voices of the skeptics became increasingly quieter. And as we face the world of \textit{real} high pressure, both NMR and high pressure communities slowly begin to show interest in each other.  
\section{Acknowledgements}

I am particularly thankful to Leonid Dubrovinsky, who offered me a safe haven to pursue my research. Without his encouragement and enthusiasm, we would not stand were we are today. The same holds true for Jan Korvink, who showed me that it is always worthwhile to look beyond the own comfort zone and find solutions where you would not expect them. Also, I want to thank my wife who, despite endless proof readings, always encouraged me, even in times of upmost discord. Finally, I want to thank the editorial board of PNMR for the chance to write about my story and my contribution to, what I think, is one of the most intriguing new research brands for the magnetic resonance community.













\section{Bibliography}

\begin{thebibliography}{59}
\providecommand{\natexlab}[1]{#1}
\providecommand{\url}[1]{\texttt{#1}}
\expandafter\ifx\csname urlstyle\endcsname\relax
  \providecommand{\doi}[1]{doi: #1}\else
  \providecommand{\doi}{doi: \begingroup \urlstyle{rm}\Url}\fi

\bibitem[Hazen(1993)]{Hazen1993}
R.~M. Hazen.
\newblock \emph{{The New Alchemists - Breaking Through the Barriers of
  High-Pressure}}.
\newblock Random House, New York, first edition, 1993.
\newblock ISBN 978-0812922752.

\bibitem[Bridgman(1952)]{Bridgman1952a}
P.~W. Bridgman.
\newblock \emph{{Physics of high pressure}}.
\newblock Dover Publications, New York, first edition, 1952.

\bibitem[Weir et~al.(1959)Weir, Lippincott, {Van Valkenburg}, and
  Bunting]{Weir1959}
C.~E. Weir, E.~R. Lippincott, A~{Van Valkenburg}, and E.~N. Bunting.
\newblock {Infrared studies in the 1- to 15-micron region to 30,000
  atmospheres}.
\newblock \emph{Journal of Research of the National Bureau of Standards Section
  A: Physics and Chemistry}, 63A\penalty0 (1):\penalty0 55, jul 1959.
\newblock \doi{10.6028/jres.063A.003}.

\bibitem[Piermarini(1993)]{Piermarini1993}
G.~J. Piermarini.
\newblock {Alvin Van Valkenburg and the diamond anvil cell}.
\newblock \emph{High Pressure Research}, 11\penalty0 (5):\penalty0 279--284,
  dec 1993.
\newblock \doi{10.1080/08957959308203156}.

\bibitem[Grochala et~al.(2007)Grochala, Hoffmann, Feng, and
  Ashcroft]{Grochala2007}
W.~Grochala, R.~Hoffmann, J.~Feng, and N.~W. Ashcroft.
\newblock {The chemical imagination at work in very tight places}.
\newblock \emph{Angewandte Chemie - International Edition}, 46\penalty0
  (20):\penalty0 3620--3642, jan 2007.
\newblock \doi{10.1002/anie.200602485}.

\bibitem[Bassett(2009)]{Bassett2009}
W.~A. Bassett.
\newblock {Diamond anvil cell, 50th birthday}.
\newblock \emph{High Pressure Research}, 29\penalty0 (2):\penalty0 163--186,
  jun 2009.
\newblock \doi{10.1080/08957950802597239}.

\bibitem[Hemley(2010)]{Hemley2010}
R.~J. Hemley.
\newblock {Percy W. Bridgman's second century}.
\newblock \emph{High Pressure Research}, 30\penalty0 (4):\penalty0 581--619,
  dec 2010.
\newblock \doi{10.1080/08957959.2010.538974}.

\bibitem[Mao et~al.(1985)Mao, Bell, and Hemley]{Mao1985}
H.~K. Mao, P.~M. Bell, and R.~J. Hemley.
\newblock {Ultrahigh pressures: Optical observations and Raman measurements of
  hydrogen and deuterium to 1.47 Mbar}.
\newblock \emph{Physical Review Letters}, 55\penalty0 (1):\penalty0 99--102,
  jul 1985.
\newblock \doi{10.1103/PhysRevLett.55.99}.

\bibitem[Dubrovinskaia et~al.(2016)Dubrovinskaia, Dubrovinsky, Solopova,
  Abakumov, Turner, Hanfland, Bykova, Bykov, Prescher, Prakapenka, Petitgirard,
  Chuvashova, Gasharova, Mathis, Ershov, Snigireva, and
  Snigirev]{Dubrovinskaia2016}
N~Dubrovinskaia, L~Dubrovinsky, N~A Solopova, A~Abakumov, S~Turner, M~Hanfland,
  E~Bykova, M~Bykov, C~Prescher, V~B Prakapenka, S~Petitgirard, I~Chuvashova,
  B~Gasharova, Y.-L. Mathis, P~Ershov, I~Snigireva, and A~Snigirev.
\newblock {Terapascal static pressure generation with ultrahigh yield strength
  nanodiamond}.
\newblock \emph{Science Advances}, 2\penalty0 (7):\penalty0
  e1600341----e1600341, jul 2016.
\newblock ISSN 2375-2548.
\newblock \doi{10.1126/sciadv.1600341}.
\newblock URL
  \url{http://advances.sciencemag.org/cgi/doi/10.1126/sciadv.1600341}.

\bibitem[Kitahara and Akasaka(2003)]{Kitahara2003}
Ryo Kitahara and Kazuyuki Akasaka.
\newblock {Close identity of a pressure-stabilized intermediate with a kinetic
  intermediate in protein folding}.
\newblock \emph{Proceedings of the National Academy of Sciences}, 100\penalty0
  (6):\penalty0 3167--3172, mar 2003.
\newblock ISSN 0027-8424.
\newblock \doi{10.1073/pnas.0630309100}.

\bibitem[Li and Akasaka(2006)]{Li2006}
Hua Li and Kazuyuki Akasaka.
\newblock {Conformational fluctuations of proteins revealed by variable
  pressure NMR}.
\newblock \emph{Biochimica et Biophysica Acta (BBA) - Proteins and Proteomics},
  1764\penalty0 (3):\penalty0 331--345, mar 2006.
\newblock ISSN 15709639.
\newblock \doi{10.1016/j.bbapap.2005.12.014}.
\newblock URL
  \url{http://linkinghub.elsevier.com/retrieve/pii/S1570963905004498}.

\bibitem[Jonas and Jonas(1994)]{Jonas1994}
J.~Jonas and A.~Jonas.
\newblock {HIGH-PRESSURE NMR SPECTROSCOPY OF PROTEINS AND MEMBRANES}.
\newblock \emph{Annu. Rev. Biophys. Biomo. Struct.}, 23:\penalty0 287--318,
  1994.
\newblock \doi{10.1146/annurev.bb.23.060194.001443}.

\bibitem[Ballard and Jonas(1997)]{Ballard1997}
L.~Ballard and J.~Jonas.
\newblock {High-Pressure NMR}.
\newblock \emph{Annual Reports on NMR Spectroscopy}, 33:\penalty0 115--150,
  1997.

\bibitem[Roche et~al.(2017)Roche, Royer, and Roumestand]{Roche2017}
Julien Roche, Catherine~A. Royer, and Christian Roumestand.
\newblock {Monitoring protein folding through high pressure NMR spectroscopy}.
\newblock \emph{Progress in Nuclear Magnetic Resonance Spectroscopy},
  102-103:\penalty0 15--31, 2017.
\newblock ISSN 00796565.
\newblock \doi{10.1016/j.pnmrs.2017.05.003}.

\bibitem[Liebermann(2011)]{Liebermann2011}
R.~C. Liebermann.
\newblock {Multi-anvil, high pressure apparatus: a half-century of development
  and progress}.
\newblock \emph{High Pressure Research}, 31\penalty0 (4):\penalty0 493--532,
  2011.
\newblock \doi{10.1080/08957959.2011.618698}.

\bibitem[Yousuf and Rajan(1982)]{Yousuf1982}
M.~Yousuf and K.~G. Rajan.
\newblock {Principle of massive support in the opposed anvil high pressure
  apparatus}.
\newblock \emph{Pramana}, 18\penalty0 (1):\penalty0 1--15, jan 1982.
\newblock \doi{10.1007/BF02846528}.

\bibitem[Yarger et~al.(1995)Yarger, Nieman, Wolf, and Marzke]{Yarger1995}
J.~L. Yarger, R.~A. Nieman, G.~H. Wolf, and R.~F. Marzke.
\newblock {High-Pressure 1H and 13C Nuclear Magnetic Resonance in a Diamond
  Anvil Cell}, 1995.

\bibitem[Lee(1989)]{Lee1989a}
S.~H. Lee.
\newblock \emph{{Diamond Anvil Cell high Pressure NMR}}.
\newblock PhD thesis, Washington University, 1989.

\bibitem[Lee et~al.(1992)Lee, Conradi, and Norberg]{Lee1992}
S.~H. Lee, M.~S. Conradi, and R.~E. Norberg.
\newblock {Improved NMR resonator for diamond anvil cells}.
\newblock \emph{Review of Scientific Instruments}, 63\penalty0 (7):\penalty0
  3674--3676, jul 1992.
\newblock ISSN 0034-6748.
\newblock \doi{10.1063/1.1143597}.
\newblock URL \url{http://aip.scitation.org/doi/10.1063/1.1143597}.

\bibitem[Pravica and Silvera(1998{\natexlab{a}})]{Pravica1998}
M.G.~G Pravica and I.~F. Silvera.
\newblock {Nuclear magnetic resonance in a diamond anvil cell at very high
  pressures}.
\newblock \emph{Review of Scientific Instruments}, 69\penalty0 (2):\penalty0
  479--484, feb 1998{\natexlab{a}}.
\newblock \doi{10.1063/1.1148686}.

\bibitem[Meier(2017)]{Meier2017b}
Thomas Meier.
\newblock {At Its Extremes: NMR at Giga-Pascal Pressures}.
\newblock In Graham Webb, editor, \emph{Annual Reports on NMR Spectroscopy},
  chapter~1, pages 1--74. Elsevier, London, 93 edition, 2017.
\newblock ISBN 9780128149133.
\newblock \doi{10.1016/bs.arnmr.2017.08.004}.

\bibitem[Pravica and Silvera(1998{\natexlab{b}})]{Pravica1998a}
M.~G. Pravica and I.~F. Silvera.
\newblock {NMR Study of Ortho-Para Conversion at High Pressure in Hydrogen}.
\newblock \emph{Physical Review Letters}, 81\penalty0 (19):\penalty0
  4180--4183, nov 1998{\natexlab{b}}.
\newblock \doi{10.1103/PhysRevLett.81.4180}.

\bibitem[Lee et~al.(1987)Lee, Luszczynski, Norberg, and Conradi]{Lee1987}
Sam-Hyeon~H. Lee, K.~Luszczynski, R.~E. Norberg, and Mark~S. Conradi.
\newblock {NMR in a diamond anvil cell}.
\newblock \emph{Review of Scientific Instruments}, 58\penalty0 (3):\penalty0
  415, 1987.
\newblock ISSN 00346748.
\newblock \doi{10.1063/1.1139246}.
\newblock URL \url{http://link.aip.org/link/RSINAK/v58/i3/p415/s1{\&}Agg=doi}.

\bibitem[Lee et~al.(1989)Lee, Conradi, and Norberg]{Lee1989}
S.~H. Lee, M.~S. Conradi, and R.~E. Norberg.
\newblock {Molecular motion in solid H2 at high pressures}.
\newblock \emph{Physical Review B}, 40\penalty0 (18):\penalty0 12492--12498,
  1989.
\newblock \doi{10.1103/PhysRevB.40.12492}.

\bibitem[Lee et~al.(2008)Lee, Charnaya, and Tien]{Lee2008}
M.~K. Lee, E.~V. Charnaya, and C.~Tien.
\newblock {Self-diffusion slowdown in liquid indium and gallium metals under
  nanoconfinement}.
\newblock \emph{Microelectronics Journal}, 39\penalty0 (3-4):\penalty0
  566--569, mar 2008.
\newblock \doi{10.1016/j.mejo.2007.07.031}.

\bibitem[Hazen et~al.(1987)Hazen, Finger, Angel, Prewitt, Ross, Mao,
  Hadidiacos, Hor, Meng, and Chu]{Hazen1987}
R.~M. Hazen, L.~W. Finger, R.~J. Angel, C.~T. Prewitt, N.~L. Ross, H.~K. Mao,
  C.~G. Hadidiacos, P.~H. Hor, R.~L. Meng, and C.~W. Chu.
\newblock {Crystallographic description of phases in the Y-Ba-Cu-O
  superconductor}.
\newblock \emph{Physical Review B}, 35\penalty0 (13):\penalty0 7238--7241,
  1987.
\newblock \doi{10.1103/PhysRevB.35.7238}.

\bibitem[Gao et~al.(1994)Gao, Xue, Chen, Xiong, Meng, Ramirez, Chu, Eggert, and
  Mao]{Gao1994}
L.~Gao, Y.~Y. Xue, F.~Chen, Q.~Xiong, R.~L. Meng, D.~Ramirez, C.~W. Chu, J.~H.
  Eggert, and H.~K. Mao.
\newblock {Superconductivity up to 164 K in HgBa2Ca(m-1)Cu(m)O(2m+2+d) (m=1, 2,
  and 3) under quasihydrostatic pressures}.
\newblock \emph{Physical Review B}, 50\penalty0 (6):\penalty0 4260--4263, aug
  1994.
\newblock \doi{10.1103/PhysRevB.50.4260}.

\bibitem[Souliou et~al.(2014)Souliou, Subedi, Song, Lin, Syassen, Keimer, and
  {Le Tacon}]{Souliou2014}
S~M Souliou, A~Subedi, Y~T Song, C~T Lin, K~Syassen, B~Keimer, and M~{Le
  Tacon}.
\newblock {Pressure-induced phase transition and superconductivity in
  YBa2Cu4O8}.
\newblock \emph{Physical Review B}, 90\penalty0 (14):\penalty0 140501, 2014.
\newblock \doi{10.1103/PhysRevB.90.140501}.

\bibitem[Tissen et~al.(1991)Tissen, Nefedova, and Emel`chenko]{Tissen1991}
V~G Tissen, M~V Nefedova, and G~A Emel`chenko.
\newblock {Disappearance of superconductivity in YBa2Cu4O8 under pressure about
  11 GPa}.
\newblock \emph{Sverkhprovodimost': Fizika, Khimiya, Tekhnika}, 23\penalty0
  (12), 1991.

\bibitem[Mito et~al.(2014)Mito, Imakyurei, Deguchi, Matsumoto, Hara, Ozaki,
  Takeya, and Takano]{Mito2014}
Masaki Mito, Takumi Imakyurei, Hiroyuki Deguchi, Kaname Matsumoto, Hiroshi
  Hara, Toshinori Ozaki, Hiroyuki Takeya, and Yoshihiko Takano.
\newblock {Effective disappearance of the meissner signal in the cuprate
  superconductor YBa2Cu4O8 under uniaxial strain}.
\newblock \emph{Journal of the Physical Society of Japan}, 83\penalty0
  (2):\penalty0 8--11, 2014.
\newblock ISSN 00319015.
\newblock \doi{10.7566/JPSJ.83.023705}.

\bibitem[Ma et~al.(2009)Ma, Eremets, Oganov, Xie, Trojan, Medvedev, Lyakhov,
  Valle, and Prakapenka]{Ma2009}
Y.~Ma, M.~I. Eremets, A.~R. Oganov, Y.~Xie, I.~A. Trojan, S.~A. Medvedev, A.~O.
  Lyakhov, M.~Valle, and V.~B. Prakapenka.
\newblock {Transparent dense sodium}.
\newblock \emph{Nature}, 458\penalty0 (7235):\penalty0 182--185, mar 2009.
\newblock ISSN 0028-0836.
\newblock \doi{10.1038/nature07786}.

\bibitem[Hanfland et~al.(2000)Hanfland, Syassen, Christensen, and
  Novikov]{Han2000}
M.~Hanfland, K.~Syassen, N.~E. Christensen, and D.~L. Novikov.
\newblock {New high-pressure phase of lithium}.
\newblock \emph{Nature}, 408\penalty0 (6809):\penalty0 174--178, nov 2000.
\newblock \doi{10.1038/35041515}.
\newblock URL \url{http://www.nature.com/doifinder/10.1038/35041515}.

\bibitem[Lacey et~al.(1999)Lacey, Subramanian, Olson, Webb, and
  Sweedler]{Lacey1999}
M.~E. Lacey, R.~Subramanian, D.~L. Olson, A.~G. Webb, and J.~V. Sweedler.
\newblock {High-Resolution NMR Spectroscopy of Sample Volumes from 1 nL to 10
  $\mu$L}.
\newblock \emph{Chemical Reviews}, 99\penalty0 (10):\penalty0 3133--3152, oct
  1999.
\newblock \doi{10.1021/cr980140f}.

\bibitem[Olson et~al.(1995)Olson, Peck, Webb, Magin, and Sweedler]{Olson1995}
D.~L. Olson, T.~L. Peck, A.~G. Webb, R.~L. Magin, and J.~V. Sweedler.
\newblock {High-Resolution Microcoil 1H-NMR for Mass-Limited, Nanoliter-Volume
  Samples}.
\newblock \emph{Science}, 270\penalty0 (5244):\penalty0 1967--1970, dec 1995.
\newblock \doi{10.1126/science.270.5244.1967}.

\bibitem[Stocker et~al.(1997)Stocker, Peck, Webb, Feng, and Magin]{Stocker1997}
J.~E. Stocker, T.~L. Peck, A.~G. Webb, M.~Feng, and R.~L. Magin.
\newblock {Nanoliter volume, high-resolution NMR microspectroscopy using a
  60-$\mu$m planar microcoil}.
\newblock \emph{IEEE Transactions on Biomedical Engineering}, 44\penalty0
  (11):\penalty0 1122--1127, 1997.
\newblock ISSN 00189294.
\newblock \doi{10.1109/10.641340}.

\bibitem[Webb(1997)]{Webb1997}
A.~G. Webb.
\newblock {Radiofrequency microcoils in magnetic resonance}.
\newblock \emph{Progress in Nuclear Magnetic Resonance Spectroscopy},
  31\penalty0 (1):\penalty0 1--42, jul 1997.
\newblock \doi{10.1016/S0079-6565(97)00004-6}.

\bibitem[Suzuki et~al.(2009)Suzuki, Yamauchi, Shimizu, Itoh, Takeshita,
  Terakura, Takagi, Tokura, Yamauchi, and Ueda]{Suzuki2009}
T.~Suzuki, I.~Yamauchi, Y.~Shimizu, M.~Itoh, N.~Takeshita, C.~Terakura,
  H.~Takagi, Y.~Tokura, Touru Yamauchi, and Yutaka Ueda.
\newblock {High-pressure V-51 NMR study of the magnetic phase diagram and
  metal-insulator transition in quasi-one-dimensional beta-Na0.33V2O5}.
\newblock \emph{Physical Review B}, 79[1] T. S\penalty0 (8):\penalty0 081101,
  feb 2009.
\newblock \doi{10.1103/PhysRevB.79.081101}.

\bibitem[Haase et~al.(2009)Haase, Goh, Meissner, Alireza, and
  Rybicki]{Haase2009}
J.~Haase, S.~K. Goh, T.~Meissner, P.~L. Alireza, and D.~Rybicki.
\newblock {High sensitivity nuclear magnetic resonance probe for anvil cell
  pressure experiments}.
\newblock \emph{Review of Scientific Instruments}, 80\penalty0 (7):\penalty0
  073905/1 -- /4, jul 2009.
\newblock \doi{10.1063/1.3183504}.

\bibitem[Meissner et~al.(2011)Meissner, Goh, Haase, Williams, and
  Littlewood]{Meissner2011}
T.~Meissner, S.~K. Goh, J.~Haase, G.~V.~M. Williams, and P.~B. Littlewood.
\newblock {High-pressure spin shifts in the pseudogap regime of superconducting
  YBa2Cu4O8 as revealed by O17 NMR}.
\newblock \emph{Physical Review B - Condensed Matter and Materials Physics},
  83\penalty0 (22):\penalty0 220517, jun 2011.
\newblock \doi{10.1103/PhysRevB.83.220517}.

\bibitem[Meier et~al.(2014)Meier, Herzig, and Haase]{Meier2014}
T.~Meier, T.~Herzig, and J.~Haase.
\newblock {Moissanite anvil cell design for Giga-Pascal nuclear magnetic
  resonance.}
\newblock \emph{The Review of scientific instruments}, 85\penalty0
  (4):\penalty0 043903, apr 2014.
\newblock \doi{10.1063/1.4870798}.

\bibitem[Meissner(2013)]{Meissner2012}
T.~Meissner.
\newblock \emph{{Exploring Nuclear Magnetic Resonance at the Highest
  Pressures}}.
\newblock PhD thesis, Leipzig University, 2013.

\bibitem[Meier(2016)]{Meier2016}
T.~Meier.
\newblock \emph{{High Sensitivity Nuclear Magnetic Resonance at Extreme
  Pressures}}.
\newblock PhD thesis, Leipzig University, 2016.

\bibitem[Meier and Haase(2014)]{Meier2014a}
T.~Meier and J.~Haase.
\newblock {High-Sensitivity Nuclear Magnetic Resonance at Giga-Pascal Pressures
  : A New Tool for Probing Electronic and Chemical Properties of Condensed
  Matter under Extreme Conditions}.
\newblock \emph{Journal of Visualized Experiments}, 92\penalty0 (92):\penalty0
  1--10, oct 2014.
\newblock \doi{10.3791/52243}.

\bibitem[Bloembergen et~al.(1948)Bloembergen, Purcell, and
  Pound]{Bloembergen1948}
N.~Bloembergen, E.~M. Purcell, and R.~V. Pound.
\newblock {Relaxation Effects in Nuclear Magnetic Resonance Absorption}.
\newblock \emph{Physical Review}, 73\penalty0 (7):\penalty0 679--712, apr 1948.
\newblock \doi{10.1103/PhysRev.73.679}.

\bibitem[Knight(1949)]{Knight1949}
W.~D. Knight.
\newblock {Nuclear Magnetic Resonance Shift in Metals}.
\newblock \emph{Physical Review}, 76\penalty0 (8):\penalty0 1259--1260, oct
  1949.
\newblock \doi{10.1103/PhysRev.76.1259.2}.

\bibitem[Korringa(1950)]{Korringa1950}
J.~Korringa.
\newblock {Nuclear magnetic relaxation and resonance line shift in metals}.
\newblock \emph{Physica}, 16\penalty0 (7-8):\penalty0 601--610, jul 1950.
\newblock \doi{10.1016/0031-8914(50)90105-4}.

\bibitem[Heitler and Teller(1936)]{Heitler1936}
W.~Heitler and E.~Teller.
\newblock {Time Effects in the Magnetic Cooling Method. I}, 1936.

\bibitem[Schr{\"{o}}der et~al.(2013)Schr{\"{o}}der, Rosenthal, Souchay,
  Petermayer, Grott, Scheidt, Gold, Scherer, and Oeckler]{Schroder2013}
T.~Schr{\"{o}}der, T.~Rosenthal, D.~Souchay, C.~Petermayer, S.~Grott, E.~W.
  Scheidt, C.~Gold, W.~Scherer, and O.~Oeckler.
\newblock {A high-pressure route to thermoelectrics with low thermal
  conductivity: The solid solution series AgInxSb1-xTe2 (x=0.1-0.6)}.
\newblock \emph{Journal of Solid State Chemistry}, 206:\penalty0 20--26, oct
  2013.
\newblock \doi{10.1016/j.jssc.2013.07.027}.

\bibitem[Welzmiller et~al.(2014)Welzmiller, Hennersdorf, Fitch, and
  Oeckler]{Welzmiller2014}
S.~Welzmiller, F.~Hennersdorf, A.~Fitch, and O.~Oeckler.
\newblock {Solid Solution Series between CdIn2Te4 and AgInTe2 Investigated by
  Resonant X-ray Scattering}.
\newblock \emph{Zeitschrift f{\{}{\"{u}}{\}}r anorganische und allgemeine
  Chemie}, 640\penalty0 (15):\penalty0 3135--3142, dec 2014.
\newblock \doi{10.1002/zaac.201400400}.

\bibitem[Range(1971)]{Range1971}
K.~J. Range.
\newblock {High-Pressure Studies on Ternary Chalcogenides with Tetrahedrally
  Coordinated Cations}.
\newblock \emph{Chemiker-Zeitung}, 95\penalty0 (1):\penalty0 3--11, 1971.

\bibitem[Meissner et~al.(2014)Meissner, Goh, Haase, Richter, Koepernik, and
  Eschrig]{Meissner2014}
T.~Meissner, S.~K. Goh, J.~Haase, M.~Richter, K.~Koepernik, and H.~Eschrig.
\newblock {Nuclear magnetic resonance at up to 10.1 GPa pressure detects an
  electronic topological transition in aluminum metal}.
\newblock \emph{Journal of Physics: Condensed Matter}, 26\penalty0
  (1):\penalty0 015501, jan 2014.
\newblock \doi{10.1088/0953-8984/26/1/015501}.

\bibitem[Meier and Haase(2015)]{Meier2015a}
T.~Meier and J.~Haase.
\newblock {Anvil cell gasket design for high pressure nuclear magnetic
  resonance experiments beyond 30 GPa}.
\newblock \emph{Review of Scientific Instruments}, 86\penalty0 (12):\penalty0
  123906, dec 2015.
\newblock \doi{10.1063/1.4939057}.

\bibitem[Spengler et~al.(2017)Spengler, While, Meissner, Wallrabe, and
  Korvink]{Spengler2017}
Nils Spengler, Peter~T. While, Markus~V. Meissner, Ulrike Wallrabe, and Jan~G.
  Korvink.
\newblock {Magnetic Lenz lenses improve the limit-of-detection in nuclear
  magnetic resonance}.
\newblock \emph{PLOS ONE}, 12\penalty0 (8):\penalty0 e0182779, 2017.
\newblock \doi{10.1371/journal.pone.0182779}.

\bibitem[Korvink and MacKinnon(2017)]{Korvink2017}
Jan~G Korvink and Neil MacKinnon.
\newblock {Microscale magnetic resonance detectors: a technology roadmap for in
  vivo metabolomics}.
\newblock \emph{arxiv}, pages 1--5, oct 2017.

\bibitem[Jouda et~al.(2017)Jouda, Kamberger, Leupold, Spengler, Hennig,
  Gruschke, and Korvink]{Jouda2017}
Mazin Jouda, Robert Kamberger, Jochen Leupold, Nils Spengler, J{\"{u}}rgen
  Hennig, Oliver Gruschke, and Jan~G. Korvink.
\newblock {A comparison of Lenz lenses and LC resonators for NMR signal
  enhancement}.
\newblock \emph{Concepts in Magnetic Resonance Part B: Magnetic Resonance
  Engineering}, page e21357, nov 2017.
\newblock \doi{10.1002/cmr.b.21357}.

\bibitem[Meier et~al.(2017)Meier, Wang, Mager, Korvink, Petigirard, and
  Dubrovinsky]{Meier2017}
T.~Meier, N.~Wang, D.~Mager, J.~G. Korvink, S.~Petigirard, and L.~Dubrovinsky.
\newblock {Magnetic Flux Tailoring through Lenz Lenses in Toroidal Diamond
  Indenter Cells : A New Pathway to High Pressure Nuclear Magnetic Resonance}.
\newblock \emph{axiv:1706.00073}, pages 1--10, 2017.

\bibitem[van Bentum et~al.(2007)van Bentum, Janssen, Kentgens, Bart, and
  Gardeniers]{VanBentum2007}
P.~J.~M. van Bentum, J.~W.~G. Janssen, A.~P.~M. Kentgens, J.~Bart, and J.~G.~E.
  Gardeniers.
\newblock {Stripline probes for nuclear magnetic resonance}.
\newblock \emph{Journal of Magnetic Resonance}, 189\penalty0 (1):\penalty0
  104--113, 2007.
\newblock \doi{10.1016/j.jmr.2007.08.019}.

\bibitem[Eremets(1996)]{Eremets1996}
M.~I. Eremets.
\newblock \emph{{High Pressure Experimental Methods}}.
\newblock Oxford University Press, Oxford New York, first edition, 1996.

\bibitem[Meier et~al.(2018)Meier, Khandarkhaeva, Petitgirard, K{\"{o}}rber,
  R{\"{o}}ssler, and Dubrovinsky]{Meier2018a}
Thomas Meier, Saiana Khandarkhaeva, Sylvain Petitgirard, Thomas K{\"{o}}rber,
  Ernst R{\"{o}}ssler, and Leonid Dubrovinsky.
\newblock {NMR close to Mega-Bar Pressures}.
\newblock \emph{in Preparation}, 2018.

\end{thebibliography}
\end{document}